\makeatletter
\@ifundefined{@parse@version@dash}{%
\def\@parse@version#1{\@parse@version@0#1}
\def\@parse@version@#1/#2/#3#4#5\@nil{%
\@parse@version@dash#1-#2-#3#4\@nil}
\def\@parse@version@dash#1-#2-#3#4#5\@nil{%
  \if\relax#2\relax\else#1\fi#2#3#4 }
}{}
\makeatother
\documentclass[aps,prx,reprint,floatfix,superscriptaddress]{revtex4-2}	
\usepackage{bm}
\usepackage{amsmath,amsfonts,amssymb}
\usepackage[normalem]{ulem}
\usepackage{bbold}
\usepackage[T1]{fontenc}
\usepackage{graphicx,color}

\definecolor{myred}{RGB}{228,26,28}
\definecolor{myorange}{RGB}{225,127,0}
\definecolor{mygreen}{RGB}{77,175,74}
\definecolor{mylila}{RGB}{152,78,163}
\def\Tr{\mathrm{Tr}}

\begin{document}
\title{Monitoring quantum Otto engines}
\author{Jeongrak Son} 
\affiliation{Department of Physics and Astronomy, Seoul National University, Seoul 08826, Republic of Korea.}
\affiliation{Center for Theoretical Physics of Complex Systems, Institute for Basic Science (IBS), Daejeon 34126, Republic of Korea.}
\author{Peter Talkner}
\email{peter.talkner@physik.uni-augsburg.de}
\affiliation{Institut f\"{u}r Physik, Universit\"{a}t Augsburg, Universit\"{a}tsstra{\ss}e 1, D-86135 Augsburg, Germany.} 
\author{Juzar Thingna}
\email{jythingna@ibs.re.kr}
\affiliation{Center for Theoretical Physics of Complex Systems, Institute for Basic Science (IBS), Daejeon 34126, Republic of Korea.}
\affiliation{Basic Science Program, Korea University of Science and Technology, Daejeon 34113, Republic of Korea.}

\date{\today}
 
\begin{abstract}
Unlike classical systems, a measurement performed on a quantum system always alters its state. In this work, the impacts of two diagnostic schemes to determine the performance of quantum Otto heat engines are compared: In one scheme, the energy of the engine's working substance is measured after each stroke (repeated measurements), and in the other one, the energies after each stroke are recorded in one or two pointer states and measured only after the completion of a prescribed number of cycles (repeated contacts). A single pointer state suffices if one is only interested in either work or heat. For joint work and heat diagnostics, two pointers are needed. These schemes are applied to Otto engines, whose working substance consists of a two-level system. Depending on the engine protocol, the duration of a single cycle may be infinite or finite. Because in the repeated contact scheme, the number of measurements is drastically reduced compared to the repeated measurement scheme, the quantum coherence after and during the contact diagnostics is much better maintained than repeated measurements that destroy any coherence at the end of each stroke. We demonstrate that maximum power, reliability, and efficiency of the engine in the presence of repeated contacts typically outperform these figures of merit of repeated measurements. Due to the improved coherence persistence, heat engines with a finite cycle duration require a larger number of cycles to reach a periodically asymptotic state. Overall, our results document the importance of taking into account the particular nature of diagnostic tools for monitoring and testing purposes but also for feedback control, both in theory and experiment.  
\end{abstract}
\maketitle
\section{Introduction}\label{I}

Heat engines have not only been important devices from a technological perspective, but they also played a significant role in establishing and examining the laws of thermodynamics~\cite{Muller07}. Remarkably, with the extension of thermodynamics from macroscopic to ever smaller systems during the past decades, the interest in the same paradigmatic model systems such as Carnot and Otto engines has revived, with efficiency and power as figures of merit. A particular aspect of small engines results from the generic randomness of their dynamics due to the coupling of the microscopically small working substance with heat baths at different temperatures~\cite{Blickle12}. Already shortly after the invention of the maser the similarity of its action principle with a heat engine was noted~\cite{Scovil59}. Apart from scattered early considerations~\cite{Alicki79,Kosloff84,Geva92}, it took about forty years until this concept gained broader interest and the idea of quantum Carnot and Otto engines was further elaborated~\cite{Bender00,Feldmann03}. Since then basically all of the  distinguishing features of quantum systems have been investigated in view of their potentially beneficial or possibly detrimental aspects for an engine, such as coherence~\cite{Scully03,Uzdin16,Francica19,Dann20,Latune21}, discreteness and spacing of energy levels~\cite{Quan05,Gelbwaser18, Halpern19}, quantum correlations and entanglement~\cite{Gelbwaser15,Wang09}, Fermi versus Bose statistics~\cite{Watanabe20}, and projective measurements as an energy source~\cite{Yi17,Ding18}. Experimental investigations of quantum heat engines were suggested in Ref.~\cite{Campisi15} and realized using trapped ions \cite{Rossnagel16, Maslennikov19}, optomechanical systems \cite{Holmes20}, nitrogen-vacancy centers \cite{Klatzow19}, and NMR techniques~\cite{Peterson19}.

So far, the monitoring of engines has found only relatively little attention, be it as a merely diagnostic tool or as a basic element of a feedback control device~\cite{Quan06}. The functioning of four stroke Otto-like engines during a single cycle was tracked in Refs.~\cite{Zheng16,Ding18,Denzler21} by projective energy measurements of the working substance immediately after the completion of each stroke. From these four measured values the energy changes during work strokes and thermalization strokes corresponding to work and heat, respectively, were characterized by their joint statistics. The reduction of the engine's working substance state with respect to the instantaneous energy basis specifying each measurement, in general alters the  subsequent dynamics and hence may influence the resulting energy differences between subsequent measurements. This total suppression of coherence is specific for the two point projective measurement scheme of work~\cite{Kurchan00,Tasaki00,Talkner07} and has often been interpreted as a deficiency~\cite{Kammerlander16,Vinjanampathy16}. On the other hand, the traditional assignment of average work done on a thermally closed system by a control parameter change as the difference of the average energies calculated after and before the forcing~\cite{Alicki79,Nieuwenhuizen02} cannot be extended to the definition of a proper random variable~\cite{Allahverdyan14,Talkner16,Perarnau17,Talkner20}. Only if the interference of the system with a particular monitoring device is specified, e.g. in terms of two projective energy measurements in the two point measurement scheme or by two subsequent interactions of the system with the same pointer state of a measurement apparatus~\cite{Talkner16}, or by any other generalized measurement scheme~\cite{Wiseman09}, work can be properly specified as a random variable at the price that its probability distribution depends on the chosen measurement procedure. This principle restriction still leaves the possibility to look for detection schemes that optimize particular features such as the accuracy or the amount of back-action on the system, to name just two possibly complementary criteria.                                        

For the diagnostics of a reciprocating engine such as an Otto engine, the monitoring of the energy of the working substance several times during a single cycle is crucial in order to be able to infer the total performed work and the consumed heat during a single, and also over many cycles. With projective energy measurements any coherence possibly built up during a work stroke~\cite{Francica19} or after a thermalization stroke~\cite{Thingna12} would be completely removed with an impact on the performance of the subsequent strokes. The coherencies of the bath have been identified as a potential boost to the performance of an engine~\cite{Scully03}, while those built up in the working substance have been found as potentially detrimental~\cite{Feldmann03,Rezek06,Plastina14,Francica19}.  

In the present paper we adapt a recently proposed method to determine the sum of an observable by means of repeated contacts~\cite{Thingna19} to find the total work and the heat supplied to an engine in a prescribed number of cycles. We compare the obtained results with those from repeated measurements and demonstrate that the repeated contact method is in general less invasive and also leads to a better performance of the engine.   

We organize the paper as follows: In Sec.~\ref{sec:engine} we introduce the general quantum Otto engine and formulate the repeated contact and repeated measurement schemes. General expressions for the joint work and heat probability density functions (PDFs) are presented for both schemes. In Sec.~\ref{PT}, a two-level system, i.e. a single qubit, is  considered as the working substance  for which general expressions for the work strokes are presented. For the heat strokes we consider perfect thermalization beyond weak system-bath coupling and alternatively finite time thermalization at weak system-bath coupling. We compare our analytic expressions of work and heat PDFs as well as the second moments resulting from the two measurement schemes for a single engine cycle. In Sec.~\ref{NR} numerical results for several cycles are presented. Finally, in Sec.~\ref{Conclusions}, we draw our conclusions.

\section{Quantum Otto engines}\label{sec:engine}
\subsection{Mode of operation}
\begin{figure}[t!]
\includegraphics[width=\columnwidth]{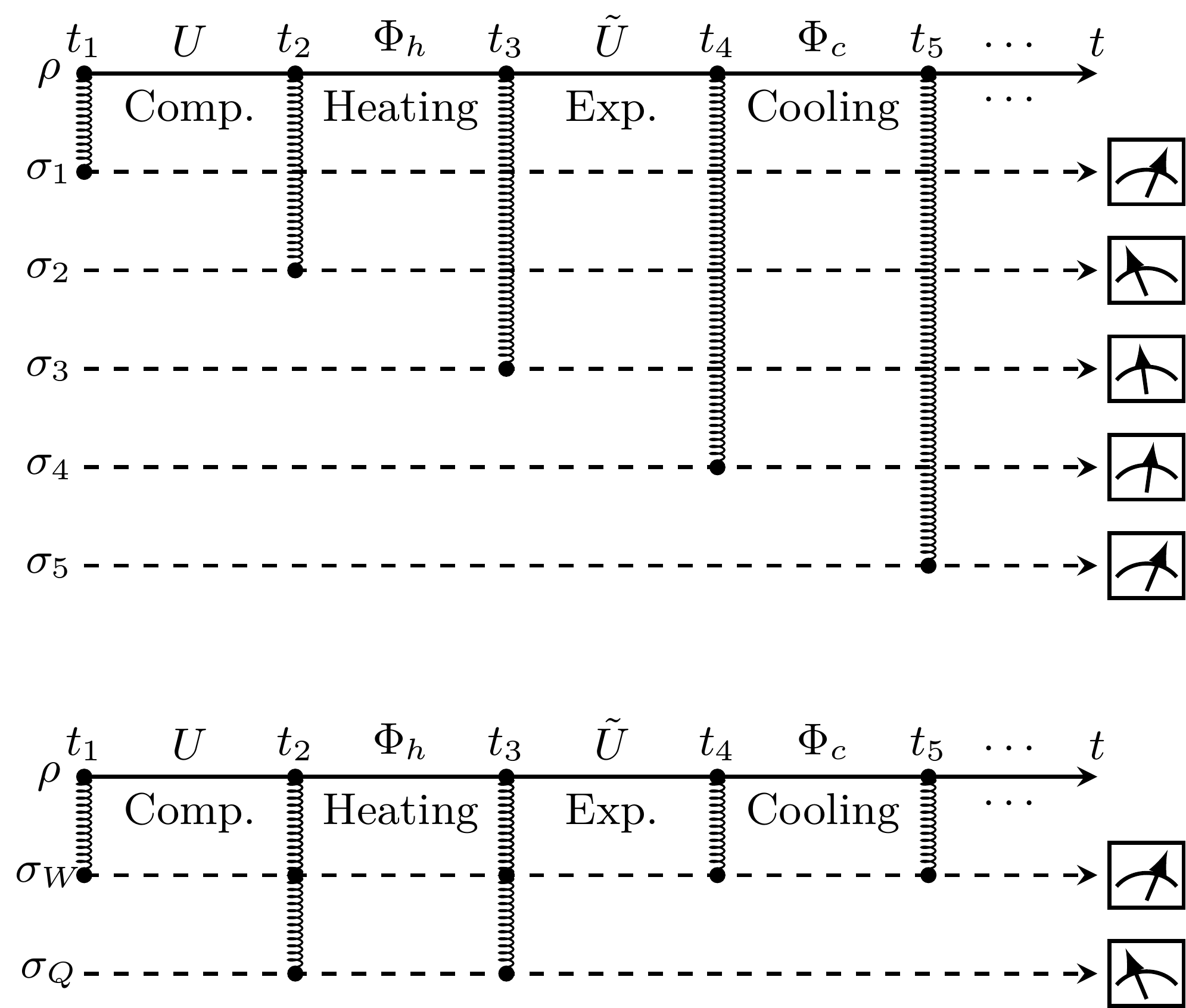}
\caption{Illustration for repeated measurements and two-pointer repeated contacts applied to quantum Otto engines running for $N$ cycles. The top panel illustrates repeated measurements with $4N$ pointers, five pointers of which are shown. Each pointer is initialized in the same state $\sigma$. The measurements are taken at the beginnings and the end of the work strokes when the working substance is thermally isolated. The results of the measurements are symbolically presented at the right margin. The registered pointer states can be used to calculate heat and work.       
The lower panel illustrates the repeated contact scheme with two pointers, the upper one for the work and the lower one for the heat. Again both pointers are initialized in the same state $\sigma$. Here the actions of the contacts on the two pointers are accumulated with appropriate signs, see the  Sec.~\ref{dmI} for the details, and read out by a projective position measurement after a prescribed number of engine cycles is completed. Between subsequent contacts the states of the two pointers do not evolve.}
\label{Fig0}
\end{figure}

For the sake of concreteness we consider an Otto engine with a quantum mechanical working substance, which later will be specified as a two-level system. The subsequent four strokes constituting a full cycle of the engine consist in compressing, heating, expanding, and cooling the working substance, see Fig.~\ref{Fig0}. The compression and expansion strokes are caused by changing a control parameter of the working substance in thermal isolation. Therefore, the energy change of the working substance during these strokes can be referred to as work. The parameter change in the compression stroke is designed such that the energy level distances of the working substance Hamiltonian increase. Hence, the time evolution of the density matrix caused by the compression is determined by a unitary operator $U$ with $\rho_2 = U \rho_1 U^\dagger$, with $\rho_k =\rho(t_k)$, $k = 1,2,\cdots $, denoting the density matrices at the times indicated in Fig.~\ref{Fig0}, whereby times $t_k$ with $k=1,\cdots,4$ specify instants within the first cycle, with $k=5,\cdots, 8$ within the second cycle, etc.. For the sake of simplicity, the expansion is assumed to be the  time-reversed protocol of the compression. Therefore, the unitary operator giving the corresponding transformation can be written as $\tilde{U} = K^{*}U^{\dagger}K$ with $K^{*}$ being the adjoint of the antiunitary time-reversal operator $K$ ($K^{*}K=KK^*=\mathbb{1}$),~\cite{Messiah},  such that $\rho_4 = \tilde{U}\rho_3\tilde{U}^{\dagger}$. The relation between $\tilde{U}$ and $U^{\dagger}$ holds as long as $K^{*}H(t)K = H(t)$ for all time $t$ during the work strokes with $H(t)$ being the time-dependent engine Hamiltonian. During the second and the fourth stroke the working substance is brought into contact with a hot and a cold heat bath, respectively, while its Hamiltonian remains unchanged. These two heat strokes are characterized by trace preserving, positive, linear  maps $\Phi_h$ and $\Phi_c$ relating the respective density matrices of the working substance at the beginnings and the ends of these strokes according to $\rho_3 = \Phi_h(\rho_2)$ and $\rho_5 = \Phi_c(\rho_4)$. Later, we shall give explicit expressions specifying the unitary operator $U$, $\tilde{U}$ and the maps $\Phi_h$ and $\Phi_c$ for the case of a two-level system as working substance.     

\subsection{Diagnostic means I: Repeated measurements}\label{dmI}                   
The central quantities characterizing the performance of an engine are the total work $W$ performed on and the total heat $Q$ supplied by the hot bath to the working substance during $N$ cycles. Both quantities follow from the sequence of energies $\mathbf{e} = (e_1,e_2, \cdots, e_{4N} )$ of the working substance taken at the beginning and the end of each work stroke when the working substance is decoupled from either heat bath, see Fig.~\ref{Fig0}  yielding
\begin{eqnarray}
W &=& \sum_{k=1}^{4N} (-1)^k e_k,\label{W}\\
Q &=&\sum_{k=1}^{4N}  \mu_k e_k \label{Q},
\end{eqnarray}      
where $\mu_{1+4l}=\mu_{4+4l}= 0$,  $\mu_{2+4l}=-1$, $\mu_{3+4l}=1$ with $l=0, \cdots, N-1$.

In the repeated  measurement scheme the energies $e_k$ are measured individually by bringing the working substance for a short time into contact with the pointer of a measurement apparatus~\cite{vonNeumann32}. This contact causes a unitary transformation of the joint state of the pointer and the working substance given by
\begin{equation}
V_k = e^{-i \kappa H_k\otimes P_k},
\end{equation}  
where $H_k$ is the working substance Hamiltonian at the start of the $k$th stroke, i.e. $H_1=H_4=H_5=\cdots=H_c$ and $H_2=H_3=\cdots=H_h$, see also Fig. \ref{Fig0}; furthermore,  $P_k$ denotes the momentum operator conjugate to the pointer coordinate of the $k$th measurement apparatus. The parameter $\kappa$ results as the product of the strength and the duration of the contact which is assumed to be so short that the free motion of the working substance appears as frozen. Above, and throughout this work we will set $\hbar=k_B=1$. Before the $k$th interaction, the working substance and the respective pointer are uncorrelated. The latter resides in a Gaussian pure state $\sigma_k$ with vanishing mean value and variance $\Sigma^2$. Its position matrix element $\sigma(x,y) = \langle x |\sigma_k| y \rangle$ is independent of $k$ and is given by 
\begin{equation}
\sigma(x,y) = \frac{1}{\sqrt{2\pi  \Sigma^2}}\exp \left [-\frac{x^{2}+y^{2}}{4 \Sigma^{2} }\right].
\label{Gpointer}
\end{equation}
If, after the contact has taken place, the pointer position is measured projectively at a position $x$, the non-normalized density matrix of the working substance becomes
\begin{eqnarray}
\phi_x(\rho)&=&\langle x | V_k\rho\otimes\sigma_k V^{\dagger}_k | x\rangle \nonumber \\
&=& \sum_{m,m^{\prime}}\mathcal{P}^m_{k}\rho\mathcal{P}^{m^{\prime}}_k\sigma_k\left(x - e^{m}_k,x- e^{m^{\prime}}_k\right),\label{single_contact}
\end{eqnarray}
where $e^m_k$ and $\mathcal{P}^m_k$ denote the eigenvalues and eigen-projectors of the Hamiltonian $H_k =\sum_m e^m_k \mathcal{P}^m_k$ and $\rho$ denotes the density matrix of the working substance before the contact. For the sake of simplicity and without restriction of generality, the pointer position is measured in units of energy such that $\kappa =1$. The subsequent measurements of the energies $e_k$ during $N$ cycles, giving rise to the pointer positions $\vec{x}_{N}=(x_1,\cdots, x_{4N})$, lead to a mapping of the initial working substance density matrix $\rho$ onto the non-normalized density matrix $\phi^{\mathrm{RM}}_{\vec{x}_N}(\rho)$ given by
\begin{equation}
\phi^{\mathrm{RM}}_{\vec{x}_{N}}(\rho) = \sum_{\vec{m}_N,\vec{m}^{\prime}_N}D^{\vec{m}_N,\vec{m}^{\prime}_N}(\rho)\prod_{k=1}^{4N}\sigma \left(x_{k}-e_{k}^{m_{k}},x_{k}-e_{k}^{m_{k}^{\prime}}\right),
\label{RM_tot}
\end{equation}
where the sum runs over all possible sequences $\vec{m}^{(\prime)}_N =(m^{(\prime)}_1,m^{(\prime)}_2,\cdots, m^{(\prime)}_{4N})$  of quantum numbers $m^{(\prime)}_k$ labelling the eigenstates of the Hamiltonian $H_k$. The operator $D^{\vec{m}_N,\vec{m}^{\prime}_N}(\rho)$ describes the action of the alternating time-evolution operators characterizing the subsequent strokes and the interrupting projections due to the energy measurements on the initial density matrix $\rho$. They are recursively given by     
\begin{equation}
D^{\vec{m}_{N+1},\vec{m}^{\prime}_{N+1}}(\rho) = S^{\mathbf{m},\mathbf{m}^{\prime}} \left(D^{\vec{m}_N,\vec{m}^{\prime}_N}(\rho)\right), 
\label{Dmap}
\end{equation}
where the sequences $\vec{m}^{(\prime)}_{N+1} = \vec{m}^{(\prime)}_N \oplus \mathbf{m}^{(\prime)}$ denote the concatenation of $\vec{m}^{(\prime)}_N$ and $\mathbf{m}^{(\prime)} = (m_{4N+1}^{(\prime)},\cdots, m_{4(N+1)}^{(\prime)})$. Furthermore, the map $S^{\mathbf{m},\mathbf{m}^{\prime}}$ acts as 
\begin{widetext}
\begin{eqnarray}
\label{Smap}
S^{\mathbf{m},\mathbf{m}^{\prime}}(\rho) &=& \Phi_c \left (\mathcal{P}^{m_{4(N+1)}}_1 \tilde{U} \mathcal{P}^{m_{4N+3}}_2  \Phi_h \left ( \mathcal{P}^{m_{4N+2}}_2 U  \mathcal{P}^{m_{4N+1}}_1 \rho  \mathcal{P}^{m_{4N+1}^\prime}_1 U^\dagger \mathcal{P}^{m_{4N+2}^\prime}_2 \right )  \mathcal{P}^{m_{4N+3}^\prime}_2  \tilde{U}^{\dagger}  \mathcal{P}^{m_{4(N+1)}^\prime}_1\right).
\label{Smm}
\end{eqnarray} 
\end{widetext}
Taking the trace of $\phi^{\mathrm{RM}}_{\vec{x}}(\rho)$ one obtains for the joint PDF $p(\vec{x})$
to find the pointers at the values given by the components of the vector $\vec{x}$ the formal expression
\begin{equation}
p(\vec{x}) =\Tr\left[ \phi^{\mathrm{RM}}_{\vec{x}}(\rho)\right].
\label{pxphi}
\end{equation}  
For Gaussian pointers as specified in Eq.~(\ref{Gpointer}) this PDF can be written as 
\begin{widetext}
\begin{equation}
p(\vec{x}) = \sum_{\vec{m},\vec{m}^\prime} D^{\vec{m},\vec{m}^\prime} e^{-\frac{1}{8 \Sigma^2} \sum_{k=1}^{4N} \left(e^{m_k}_k - e^{m_k^\prime}_k\right)^2} \prod_{k=1}^{4N} g_{\Sigma^2} \left (x_k - \frac{1}{2} \left (e^{m_k}_k + e^{m_k^\prime}_k\right) \right ),
\label{px}
\end{equation}
\end{widetext}
where $g_{\Sigma^2}(x) = (2 \pi \Sigma^2)^{-1/2} \exp[-x^2/(2 \Sigma^2)]$ denotes a Gaussian PDF with vanishing mean value and variance $\Sigma^2$. The coefficient  $D^{\vec{m},\vec{m}^{\prime}}$ is defined as
\begin{equation}
D^{\vec{m},\vec{m}^{\prime}} = \Tr[D^{\vec{m},\vec{m}^{\prime}}(\rho)].
\label{DmmDmm}
\end{equation}
In order not to overburden the notation we omitted the index $N$ in the vectors $\vec{m}^{(\prime)}$ and $\vec{x}$.   
  
Identifying the components $x_k$ as the measured energies, one obtains for the joint probability of work and heat, $P^{\mathrm{RM}}(W,Q)$, the expression 
\begin{eqnarray}
 P^{\mathrm{RM}}(W,Q) &=& \int d^{4N}x\, \delta \left (W-\sum_{k=1}^{4N}(-1)^k  x_k \right ) \nonumber \\
 &&\times \delta \left (Q- \sum_{k=1}^{4N} \mu_k x_k \right ) p(\vec{x}).  
\label{PRMWQx}  
\end{eqnarray}
Going to the characteristic function $G^{\mathrm{RM}}(u,v) = \int dW dQ e^{iu W} e^{iv Q} P^{\mathrm{RM}}(W,Q)$ one can 
perform the $4N$-fold $x$-integration to find
\begin{eqnarray}
 G^{\mathrm{RM}}(u,v) &= 
\sum_{\vec{m},\vec{m}^\prime} D^{\vec{m},\vec{\tilde{m}}^{\prime}} e^{-\frac{1}{8 \Sigma^2} \sum_{k=1}^{4N} \left(e^{m_k}_k - e^{m_k^\prime}_k\right)^2}  \\
&\quad \times e^{iu \mathcal{W}^{\vec{m},\vec{m}^\prime} } e^{iv \mathcal{Q}^{\vec{m},\vec{m}^\prime} } e^{-N \Sigma^2 (2u^2 -2uv +v^2)},\nonumber 
\label{GRMWQ}    
\end{eqnarray}
where 
\begin{eqnarray}
\mathcal{W}^{\vec{m},\vec{m}^\prime} &= \frac{1}{2}\sum_{k=1}^{4N} (-1)^k \left ( e^{m_k}_k + e^{m_k^\prime}_k \right ), \label{Wmm}\\
 \mathcal{Q}^{\vec{m},\vec{m}^\prime} &= \frac{1}{2}\sum_{k=1}^{4N} \mu_k \left ( e^{m_k}_k + e^{m_k^\prime}_k \right ), \label{Qmm}
\end{eqnarray}
with $\mu_k$ as defined below Eq.~(\ref{Q}). This characteristic function corresponds to a linear superposition of Gaussian PDFs $g_{\mathcal{M}}(W-\mathcal{W}^{\vec{m},\vec{m}^\prime},Q-\mathcal{Q}^{\vec{m},\vec{m}^\prime})$  with mean values $ \langle W \rangle =\mathcal{W}^{\vec{m},\vec{m}^\prime}$ and $\langle Q \rangle = \mathcal{Q}^{\vec{m},\vec{m}^\prime}$ and with the covariance matrix $\mathcal{M}$ given by
\begin{equation}
\mathcal{M} = \left ( \begin{array}{cc}
\langle (\delta W)^2 \rangle & \langle \delta W \delta Q \rangle \\
\langle \delta W \delta Q \rangle & \langle (\delta Q)^2 \rangle
\end{array}
\right)
= 2 N \Sigma^2 \left (\begin{array}{cc}
 2&-1\\
-1& 1
\end{array}
\right),
\end{equation}
with $\delta X = X - \langle X\rangle$ ($X=W,Q$). Hence, the PDF is given by
\begin{eqnarray}
P^{\mathrm{RM}}(W,Q) &= \sum_{\vec{m},\vec{m}^\prime} D^{\vec{m},\vec{m}^\prime} e^{-\frac{1}{8 \Sigma^2}{\sum_{k=1}^{4N}} (e^{m_k}_k-e^{m_k\prime}_k)^2 } \nonumber \\ 
& \quad \times g_\mathcal{M}(W-\mathcal{W}^{\vec{m},\vec{m}^\prime},Q-\mathcal{Q}^{\vec{m},\vec{m}^\prime}).
\label{PRMWQ}
\end{eqnarray}
All diagonal coefficients with $\vec{m} = \vec{m}^\prime$ are non-negative whereas nondiagonal contributions with $\vec{m} \neq \vec{m}^\prime$ may assume complex values. Yet, by construction, the total sum is guaranteed to be always non-negative. Moreover, the nondiagonal contributions are suppressed by exponential factors $ \exp[-\sum_{k=1}^{4N} \left(e^{m_k}_k - e^{m_k^\prime}_k\right)^2/(8 \Sigma^2)]$. When the pointer state variance $\Sigma^2$ is sufficiently small compared to the minimum squared energy level distance of either Hamiltonian $H_c$ or $H_h$, all non-diagonal elements are negligibly small and hence only Gaussians having a maximum at $\mathcal{W}^{\vec{m},\vec{m}}, \mathcal{Q}^{\vec{m},\vec{m}}$ contribute. Due to the $N$-dependence of the covariance matrix the widths of each of the contributions grow proportionally to the square root of the number of cycles. The marginal work and heat PDFs result as
\begin{widetext}
\begin{eqnarray}
P^{\mathrm{RM}}(W) &= \sum_{\vec{m},\vec{m}^\prime} D^{\vec{m},\vec{m}^\prime} e^{-\frac{1}{8 \Sigma^2}{\sum_{k=1}^{4N}} (e^{m_k}_k-e^{m_k\prime}_k)^2 } g_{4N\Sigma^2}(W-\mathcal{W}^{\vec{m},\vec{m}^\prime}), \label{PRMW}\\
P^{\mathrm{RM}}(Q) &= \sum_{\vec{m},\vec{m}^\prime} D^{\vec{m},\vec{m}^\prime} e^{-\frac{1}{8 \Sigma^2}{\sum_{k=1}^{4N}} (e^{m_k}_k-e^{m_k\prime}_k)^2 } g_{2N\Sigma^2}(Q-\mathcal{Q}^{\vec{m},\vec{m}^\prime}). \label{PRMQ}
\end{eqnarray}
\end{widetext}
The more pronounced broadening of the Gaussian contributions in the marginal work PDF compared to the marginal heat PDF is caused by the twofold number of energy measurements required for the work compared to the heat.

\subsection{Diagnostic means II: Repeated contacts}\label{dmII}
Instead of registering each of the $4N$ energy values in a separate pointer state one may follow the idea of Ref.~\cite{Thingna20} and subsequently transfer the actual energy value at the beginning of each stroke with the appropriate sign according to the Eqs.~(\ref{W}) and~(\ref{Q}) to two pointer states, one for work and the other one for heat. The transfer of information from the working substance to the pointer states is mediated in a similar way as for repeated measurements by a sufficiently short interaction causing a unitary transformation $V_k$ of the form
\begin{equation}
V_k = e^{i \kappa H_k \otimes ((-1)^k P_W + \mu_k P_Q)},
\label{VWQk}
\end{equation}
acting on the joint Hilbert space of the working substance and the two pointer states. Here $P_W$ and $P_Q$ denote the momentum operators that are canonically conjugate to the position operators of the work and heat pointers, respectively. In contrast to the repeated measurement scheme, the pointers are only read out by a projective position measurements after the completion of $N$ cycles. Again, the pointers are assumed to be gauged in units of energy. Between two subsequent contacts the working substance does not interact with the two pointers but their states remain correlated as a result of the previous contacts. To simplify the further analysis we assume that the pointers have no own dynamics, i.e. they change their states only due to the interactions as specified by the unitary operators defined in Eq.~(\ref{VWQk}). The subsequent strokes and contacts, resulting 
after the completion of $N$ cycles in pointer state positions at specific values $W$ and $Q$,  lead to a non-normalised reduced density matrix of the working substance, which is given by            
\begin{eqnarray}
\phi^{\mathrm{RC}}_{W,Q}(\rho) &=& \sum_{\vec{m},\vec{m}^\prime} D^{\vec{m},\vec{m}^\prime}(\rho) \sigma(W-\mathcal{W}^{\vec{m}}, W- \mathcal{W}^{\vec{m}^\prime})  \nonumber \\ 
&&\times\sigma(Q-\mathcal{Q}^{\vec{m}}, Q- \mathcal{Q}^{\vec{m}^\prime}),
\label{RCWQ_red}
\end{eqnarray} 
where $\rho$ denotes the initial density matrix of the working substance and individual work and heat outcomes are denoted as 
\begin{eqnarray}
  \mathcal{W}^{\vec{m}} &=& \sum_{k=1}^{4N} (-1)^k e^{m_k}_k,\label{WRC}\\
\mathcal{Q}^{\vec{m}} &=& \sum_{k=1}^{4N} \mu_k e^{m_k}_k,\label{QRC}
\end{eqnarray}   
with $\mu_k$ as defined below Eq.~(\ref{Q}). The joint PDF $P^{\mathrm{RC}}(W,Q)$ of finding these work and heat values is given by the trace of $\phi^{\mathrm{RC}}_{W,Q}(\rho)$ yielding
\begin{widetext}
\begin{equation}
P^{\mathrm{RC}}(W,Q) = \sum_{\vec{m},\vec{m}^\prime} D^{\vec{m},\vec{m}^\prime}  e^{-\frac{1}{8 \Sigma^2} (\mathcal{W}^{\vec{m}} - \mathcal{W}^{\vec{m}\prime})^2} e^{-\frac{1}{8 \Sigma^2} (\mathcal{Q}^{\vec{m}} - \mathcal{Q}^{\vec{m}\prime})^2} g_{\Sigma^2}(W-\mathcal{W}^{\vec{m},\vec{m}^\prime}) g_{\Sigma^2}(Q-\mathcal{Q}^{\vec{m},\vec{m}^\prime}).
\label{PRCWQ} \end{equation}  
\end{widetext}
As for repeated measurements it is a linear superposition of Gaussian PDFs with weights $D^{\vec{m},\vec{m}^\prime} $, however, with the following differences: 
\begin{enumerate}
\item for repeated contacts the contributing Gaussians for work and heat factorize while for repeated measurements they are correlated; 
\item the widths of the contributing Gaussians are independent of $N$ given by $\Sigma$ whereas those for repeated measurements are proportional to $\sqrt{N} \Sigma$; 
\item for repeated contacts nondiagonal contributions are exponentially suppressed by rates  $[(\mathcal{W}^{\vec{m}} - \mathcal{W}^{\vec{m}^\prime}  )^2+ (\mathcal{Q}^{\vec{m}} - \mathcal{Q}^{\vec{m}^\prime}  )^2]/(8 \Sigma^2)$ while for repeated measurements the suppression is determined by rates of the form $\sum_{k=1}^{4N} (e^{m_k}_k-  e^{m_k^\prime}_k )^2/(8 \Sigma^2)$. 
\end{enumerate}
Therefore, for repeated contacts only contributions with different sequences $\vec{m}$ and $\vec{m}^\prime$ leading to different values of work and heat are penalized while in the repeated measurement approach deviations of individual components lead to a suppression of coherences due to the monitoring.     

For repeated contacts with a non-selective heat measurement the marginal PDF $P^{\mathrm{RC,2P}}(W)$ defined as
\begin{widetext}
\begin{equation}
 P^{\mathrm{RC,2P}}(W)= \int dQ  P^{\mathrm{RC}}(W,Q) = \sum_{\vec{m},\vec{m}^\prime} D^{\vec{m},\vec{m}^\prime}  e^{-\frac{1}{8 \Sigma^2} (\mathcal{W}^{\vec{m}} - \mathcal{W}^{\vec{m}\prime})^2} e^{-\frac{1}{8 \Sigma^2} (\mathcal{Q}^{\vec{m}} - \mathcal{Q}^{\vec{m}\prime})^2}  g_{\Sigma^2}(W-\mathcal{W}^{\vec{m},\vec{m}^\prime}) 
\label{PRC2P}
\end{equation}  
\end{widetext}
differs by the extra factor of $\exp[-(\mathcal{Q}^{\vec{m}} -\mathcal{Q}^{\vec{m}\prime})^2/(8 \Sigma^2)]$ from the PDF resulting from a mere work measurement performed with a single pointer state. In the latter case one obtains 
\begin{eqnarray}
 P^{\mathrm{RC,1P}}(W)&=& 
 \sum_{\vec{m},\vec{m}^\prime} D^{\vec{m},\vec{m}^\prime}  e^{-\frac{1}{8 \Sigma^2} (\mathcal{W}^{\vec{m}} - \mathcal{W}^{\vec{m}\prime})^2}  \nonumber \\
 && \times g_{\Sigma^2}(W-\mathcal{W}^{\vec{m},\vec{m}^\prime}).
\label{PRC1P}
\end{eqnarray}
For repeated contacts the same type of contextuality, i.e. the dependence of the observed results on the presence of a non-selective measurement, also holds for the marginal heat PDF. In both cases it is more pronounced for rather imprecise registration, i.e. for pointer states with relatively large variances $\Sigma^2$. For repeated measurements the joint work and heat PDF [see Eq. (\ref{PRMWQ})] results according to Eq. (\ref{PRMWQx}) in a classical way from the joint PDF $p(\vec{x})$ of all individual energies. Therefore repeated measurements do not present this particular contextuality feature.         

Once the work and heat PDFs are known for a specified diagnostic method, the performance of the engine can be characterized by its efficiency $\eta$ and reliability $R$ which are defined as
\begin{eqnarray}
\eta &=& -\frac{\langle W\rangle}{\langle Q\rangle},\\
R &=& -\frac{\langle W\rangle}{\sqrt{\langle W^{2}\rangle - \langle W\rangle^{2}}},\label{eq:Rel}
\end{eqnarray}
where $\langle \cdot \rangle$ denotes the average of the indicated quantity with respect to the PDF of the considered diagnostic method. Higher order efficiencies as introduced in Ref.~\cite{Ito19} can be obtained from higher order moments of work and heat. Also a fluctuating efficiency, $W/Q$, can in principle be determined~\cite{Verley14}, but will not be considered here.  

\section{Two-level system as working substance}\label{PT}
\subsection{Work strokes}
The general formulation outlined in Sec.~\ref{sec:engine} can be applied to any working substance undergoing adiabatic or non-adiabatic protocols. For a numerically feasible but yet representative example, we will use in this and subsequent sections a two-level system as the working substance and allow for non-adiabatic work strokes, i.e. strokes comprising  shifts of the instantaneous energy eigenstates and transitions between them during the unitary evolution. Without specifying the time-dependent Hamiltonian in detail, the system Hamiltonians $H_c$ and $H_h$ before and after the first work stroke, respectively, are written as
\begin{eqnarray}
H_{c} &=& \epsilon_{c}\left(|+_c\rangle\langle+_c| - |-_c\rangle\langle-_c|\right),\nonumber\\
H_{h} &=& \epsilon_{h}\left(|+_h\rangle\langle+_h| - |-_h\rangle\langle-_h|\right),
\label{Hif}
\end{eqnarray}
where $|-_u\rangle$ and  $|+_u\rangle$ denote the ground and the excited states, respectively, of the Hamiltonian $H_u$ with $u=h,c$. The most general form of the unitary evolution, during the work stroke, then reads
\begin{eqnarray}
U =& \sqrt{1-\alpha}\left(e^{-i\varphi}|+_h\rangle\langle+_c| + e^{i\varphi}|-_h\rangle\langle-_c|\right)\nonumber\\
& -\sqrt{\alpha}\Big (|+_h\rangle\langle-_c| -|-_h\rangle\langle+_c|\Big),\label{UM}\\
\tilde{U} =& \sqrt{1-\alpha}\left(e^{-i\varphi}|+_c\rangle\langle+_h| + e^{i\varphi}|-_c\rangle\langle-_h|\right)\nonumber\\
& +\sqrt{\alpha}\Big (|+_c\rangle\langle-_h| -|-_c\rangle\langle+_h|\Big).\label{UtildeM}
\end{eqnarray}
Above $\tilde{U}\equiv C^{*}U^{\dagger}C $ is the evolution operator for the time-reversed protocol with the antiunitary operator $K$ chosen to be the complex conjugation operator $C$. Here $\alpha \in [0,1]$ is the transition probability between the ground and the excited states and $\varphi$ is the corresponding phase (both parameters depend on the stroke protocol and in particular on its duration). The limiting case $\alpha =0$ refers to an adiabatic process while $\alpha =1$ corresponds to a swap. For example, a Landau-Zener protocol is described by the time-dependent Hamiltonian
\begin{equation}
H(t) = -\frac{vt}{2}\sigma_{z} -\epsilon_{c}\sigma_{x}\:,
\label{Shevchenko10}
\end{equation}
 where the velocity of the drive, $v=4\sqrt{\epsilon_h^2-\epsilon_c^2}/T_1$, is chosen such that the eigenvalues of the Hamiltonian change from $\pm \epsilon_c$ at $t=0$ to $\pm \epsilon_h$ at $t=T_1/2$, the time corresponding to the duration of a work stroke. The parameters specifying the unitary operator $U$ are given in Ref.~\cite{Shevchenko10}  by  
\begin{eqnarray}
\alpha &=& e^{-2\pi\delta},\quad \delta = \frac{\epsilon_{c}T_{1}}{4\sqrt{(\frac{\epsilon_{h}}{\epsilon_{c}})^{2}-1}}, \nonumber \\
\varphi &=& \frac{\pi}{4} - \delta(\log{\delta}-1) - \mathrm{arg}\Gamma(1-i\delta).\label{phi}
\end{eqnarray}
where $\Gamma(z)$ denotes the Gamma function. For other protocols characterized by a single parameter see Ref.~\cite{Barnes12} and for more general cases Ref.~\cite{ThingnaPRB14,Bandyopadhyay21}.

In the first instance we shall not refer to a particular work protocol and use the general form of a unitary time evolution $U$ as specified by Eq.~(\ref{UM}). Before we do so, we consider  the heating and cooling strokes of the two-level working substance that are achieved by a coupling to the heat reservoirs at temperatures $T_h$ and $T_c$, respectively. Depending on the duration and the strength of the respective contacts perfect or imperfect thermalization of the two-level system can be reached.

\subsection{Perfect thermalization}
Provided that the contact between the working substance and the heat bath is sufficiently long, the finally reached reduced state of the two-level system is always the same, dependent on the temperature of the heat bath and the strength of the interaction but independent of  its state $\rho$ at the beginning of the contact. Hence, the operation $\Phi_u$, $u=h,c$  describing a perfect thermalization stroke is given by the projection onto the final state $\rho_\beta$ and can be written as
\begin{equation}\label{E:1}
\Phi_u (\rho) = \rho_u\Tr[\rho].
\end{equation}  
Only for weak coupling $\rho_u$, $u=h,c$ coincides with the Gibbs state $\rho_u \propto \exp[-\beta_u H_u]$ defined by the Hamiltonian of the two-level system at the inverse temperature $\beta_u$. While these states lack coherence the latter may be found in generalized Gibbs states resulting from the partial trace of the total system's Gibbs state with respect to the bath~\cite{Talkner20,Thingna12}. For a perturbative treatment in the coupling strength up to second order see~Append.~\ref{Append_CPT}.

In the asymptotic limit of an initially sharp pointer state with $\Sigma=0$ the joint work and heat PDFs characterizing a single cycle are identical for repeated measurements and repeated contacts (two-pointer scheme), i.e., $P^{\mathrm{RM}}_1(W,Q) = P^{\mathrm{RC}}_1(W,Q)$. The first two moments can be expressed as
\begin{widetext}
\begin{eqnarray}
\langle W\rangle &=& \mathcal{A}_{c,h} \, \epsilon_{h} + \mathcal{A}_{h,c} \, \epsilon_{c}, \quad \quad
\langle W^{2}\rangle = 2\mathcal{B}_{c,h}\left(\epsilon_{c}^{2}+\epsilon_{h}^{2}\right) 
+ \left[\frac{1-\mathcal{B}_{c,h}}{1-2\alpha} -(1-2\alpha)\left(1+\mathcal{B}_{c,h}\right)\right]\epsilon_{c}\epsilon_{h},\label{PT_woutput}\\ 
\langle Q\rangle &=& -\mathcal{A}_{c,h}\,\epsilon_{h}, \quad \quad \quad \quad \quad ~~
\langle Q^{2}\rangle = 2\mathcal{B}_{c,h} \,\epsilon_{h}^{2},\label{PT_Q}\\
\langle WQ \rangle &=&  -2\mathcal{B}_{c,h} \,\epsilon_{h}^{2} - \frac{1}{2}\left[\frac{1-\mathcal{B}_{c,h}}{1-2\alpha} -(1-2\alpha)\left(1+\mathcal{B}_{c,h}\right)\right]\epsilon_{c}\epsilon_{h},\label{PT_WQ}
\end{eqnarray}
\end{widetext}
with $\mathcal{A}_{c,h} = 2\left(\alpha+d_{c}-d_{h}-2\alpha d_{c}\right)$ and $\mathcal{B}_{c,h}=1-(1-2\alpha)(1-2d_{c})(1-2d_{h})$. Here, $d_c$ and $d_h$ denote the populations of the exited states of the cold and the hot density matrices, respectively, see the Eqs.~(\ref{rhoth}) and (\ref{d}) in Append.~\ref{Append_CPT}. Even when the Gaussian pointer has a finite width $\Sigma\neq0$, the differences between repeated measurements and contacts in the average values of work and heat are of the order $\exp[-\epsilon^{2}_{c(h)}/(2\Sigma^{2})]$ which could be extremely small for a precise measurement apparatus, i.e. if $\epsilon_{c(h)} < \Sigma$. On the other hand, the second moments differ by $3\Sigma^{2} + \mathcal{O}(\exp[-\epsilon_{c(h)}^{2}/(2\Sigma^{2})])$ which is dominated by the $3\Sigma^2$ term arising due to the difference in the number of measurements. Accordingly, also the reliability [see Eq.~(\ref{eq:Rel})] differs for repeated contacts and repeated measurements. Overall, however, even though the two measurement schemes are quite different they do not show major significant differences for a perfectly thermalizing quantum Otto engine.

\subsection{Imperfect thermalization} 
\begin{figure}[t!]
\includegraphics[width=\columnwidth]{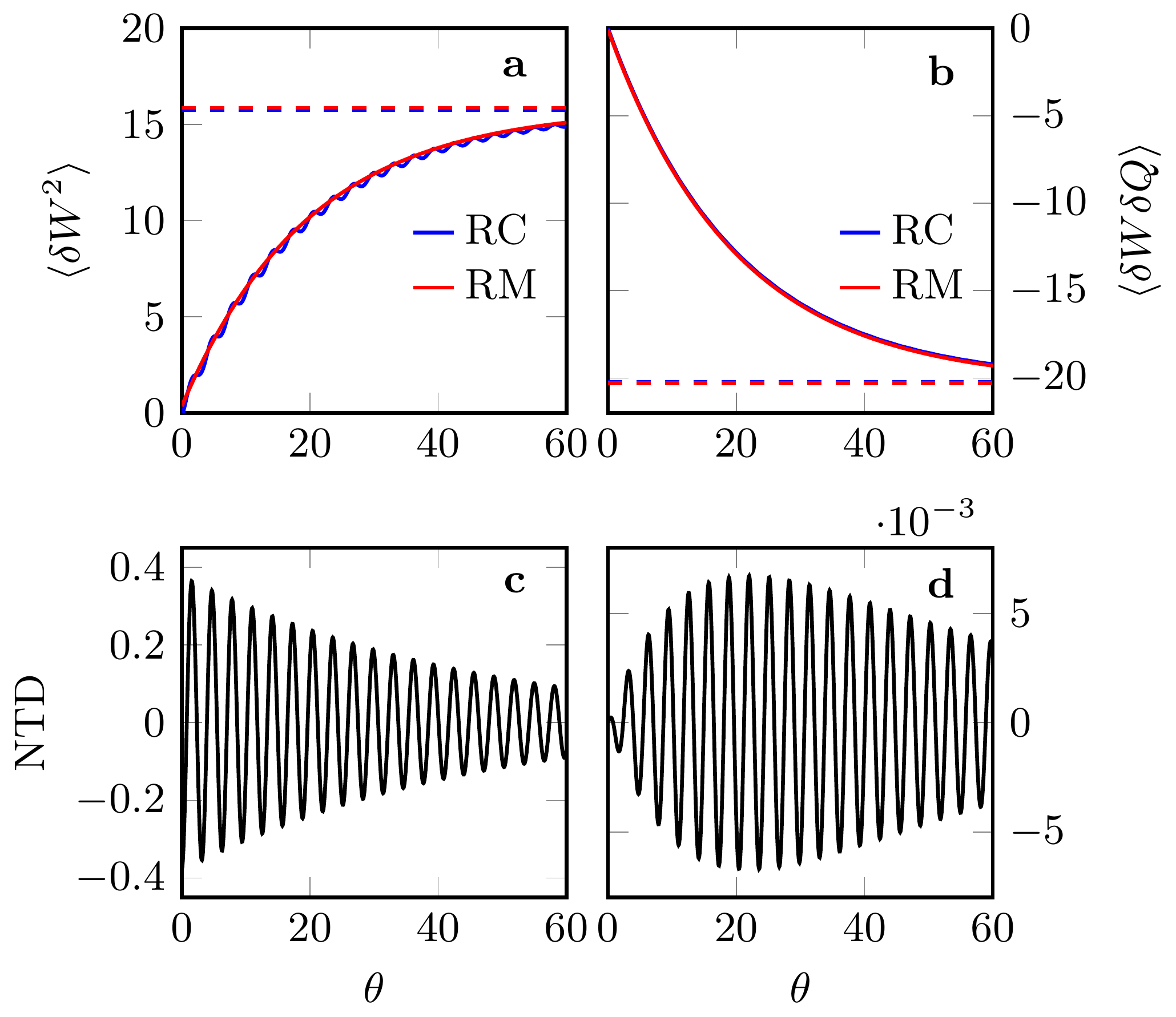}
\caption{The variance of work $\langle \delta W^2\rangle = \langle W^2 \rangle - \langle W \rangle^2$ and the covariance of work and heat, $\langle \delta W \delta Q \rangle = \langle W Q \rangle - \langle W \rangle \langle Q \rangle$,  for repeated contacts and repeated measurements for a single cycle as a function of the dimensionless time of contact with the heat baths, $\theta = \epsilon_h \tau_h=\epsilon_c \tau_c $ are drawn in panels {\bf a} and {\bf b} respectively. The dashed horizontal lines represent the asymptotic values in the perfect thermalization limit $\theta \rightarrow \infty$. In both of the plots, the differences, which are almost imperceptible in the panels {\bf a} and {\bf b}, can be divided into two parts. The first part is the constant shift proportional to $\Sigma^2$ in Eqs. (\ref{W2RCIT}) and (\ref{WQRMIT}). The remaining nontrivial differences (NTD), which are plotted in panels {\bf c} (for $\langle \delta W^2\rangle$) and {\bf d} (for $\langle \delta W \delta Q \rangle$), result from the terms proportional to $\cos (2(\theta +\varphi))$ in the Eqs.~(\ref{WRCIT}) and (\ref{W2RCIT}). The engine parameters are as follows:   The initial half energy gap $\epsilon_{c} = 1$, the final half energy gap $\epsilon_{h} = 3.7$, and the Gaussian width of the measurement is $\Sigma = 0.2$. The work strokes are characterized by $\alpha= 0.1$ and $\varphi=0$. The imperfect thermalization strokes follow a master equation dynamics which asymptotically approaches Gibbs states at the temperatures  $\beta_{c}=0.25$, $\beta_{h}=0.025$ at the relaxation rate $\gamma = 0.025$. The engine is started in the invariant state $\rho^*$ [Eq.~(\ref{rFr})] of the unobserved cycle dynamics given by the Eq.~(\ref{Fr}).}     
\label{fig_cov}
\end{figure}

Perfect thermalization requires long times during which the working substance is coupled to one of the two heat baths, leading to long cycle periods and correspondingly small power output of the engine. The dynamics of the relaxation process in general is very complicated. A substantial simplification occurs only at week coupling in the Markovian limit in which the dynamics is governed by a Lindblad equation       
\begin{equation}
\dot{\rho_u}(t) = -i[H_u,\rho_u] + \Gamma_u^+\mathcal{D}_u^{+}[\rho_u]+ \Gamma_u^-\mathcal{D}_u^{-}[\rho_u],\label{Lindblad}
\end{equation}
with the subscript $u=c,h$, the dissipators $\mathcal{D}_u^{\pm}[\rho_u] = 2L^{\mp}_u\rho_uL^{\pm}_u-\{L^{\pm}_uL^{\mp}_u,\rho_u\}$, the rates $\Gamma_u^{\pm} = \gamma\epsilon_u/(1+\exp[\mp2\beta_u\epsilon_u])$, and the Lindblad jump operators $L^{\pm}_u=|\pm_u\rangle\langle\mp_u|$. The Hamiltonians $H_u$ are given by Eq. (\ref{Hif}). Even though the stationary solutions of the master equations for the hot and the cold heat baths are given by the Gibbs states, corresponding to the Hamiltonians $H_u$  at the respective temperatures $\beta_u$ and hence do not possess any coherence, the solutions of the master equations at any finite time in general do contain coherences. This leads to differences between repeated measurements and repeated contacts even in the limit of infinite precision ($\Sigma \rightarrow 0$). For a single cycle, starting at some initial state $\rho=d|+_\mathrm{c}\rangle\langle+_\mathrm{c}| + (1-d)|-_\mathrm{c}\rangle\langle-_\mathrm{c}|+[q|+_\mathrm{c}\rangle\langle-_\mathrm{c}| + \mathrm{h.c.}]$ one finds the following expressions for the first two moments of work and heat for a small pointer state variance including leading order $\mathcal{O}(\Sigma^2)$ corrections:
\begin{widetext}
\begin{eqnarray}
\langle W\rangle^{\mathrm{RM}} &=& 2\alpha(1-\alpha)\left[\epsilon_{h}\tanh\left(\beta_{h}\epsilon_{h}\right)(1-e^{-2\gamma\theta}) +\epsilon_{c}(1-2d)(1+e^{-2\gamma\theta})\right]\nonumber\\
&&+\sum_{j=0,1}\left[(-1)^{j}\epsilon_{h}+\epsilon_{c}\right]\left[1+(-1)^{j}\left(2\alpha-1\right)\right]^{2} \left(\frac{e^{(-1)^{j}\beta_{h}\epsilon_{h}}}{e^{\beta_{h}\epsilon_{h}}+e^{-\beta_{h}\epsilon_{h}}}-d\right)\left(1-e^{-2\gamma\theta}\right),  \label{WRMIT}\\
\langle W\rangle^{\mathrm{RC}} &=& \langle W\rangle^{\mathrm{RM}} -4\epsilon_{c}\alpha(1-\alpha)(1-2d)e^{-\gamma\theta}\cos\left(2(\theta+\varphi)\right) , \label{WRCIT}\\
\langle W^{2}\rangle^{\mathrm{RM}} &=& 4\left(\epsilon_{h}^{2}+\epsilon_{c}^{2}\right) 
 \bigg \{ (\alpha+d-2\alpha d)+\left[ (1-2d )(1-2\alpha)  -2\alpha^{2}d\right] \frac{e^{-\beta_{h}\epsilon_{h}}}{e^{\beta_{h}\epsilon_{h}}+e^{-\beta_{h}\epsilon_{h}}}\bigg \}  \left(1-e^{-2\gamma\theta}\right)\nonumber\\
&& + 8\epsilon_{h}\epsilon_{c}\bigg \{ \alpha^{2}(1-2d)-(1-2\alpha)d  -\left[(1-2d)(1-2\alpha)-2\alpha^{2}d\right]\frac{e^{-\beta_{h}\epsilon_{h}}}{e^{\beta_{h}\epsilon_{h}}+e^{-\beta_{h}\epsilon_{h}}}\bigg \}\left(1-e^{-2\gamma\theta}\right)\nonumber\\
&&+ 4\epsilon_{c}^{2}\alpha(1-\alpha)e^{-2\gamma\theta} +4\Sigma^{2},
\label{W2RMIT}\\
\langle W^{2}\rangle^{\mathrm{RC}} &=& \langle W^{2}\rangle^{\mathrm{RM}} -8\epsilon_{c}^{2}\alpha(1-\alpha)e^{-\gamma\theta}\cos\left(2(\theta+\varphi)\right)  -3\Sigma^{2},
\label{W2RCIT}\\
\langle WQ\rangle^{\mathrm{RC}}&=& -4\epsilon_{h}^{2} 
 \bigg \{ (\alpha+d-2\alpha d)+\left[ (1-2d )(1-2\alpha)  -2\alpha^{2}d\right] \frac{e^{-\beta_{h}\epsilon_{h}}}{e^{\beta_{h}\epsilon_{h}}+e^{-\beta_{h}\epsilon_{h}}}\bigg \}  \left(1-e^{-2\gamma\theta}\right)\nonumber\\
&& - 4\epsilon_{h}\epsilon_{c}\bigg \{ \alpha^{2}(1-2d)-(1-2\alpha)d -\left[(1-2d)(1-2\alpha)-2\alpha^{2}d\right]\frac{e^{-\beta_{h}\epsilon_{h}}}{e^{\beta_{h}\epsilon_{h}}+e^{-\beta_{h}\epsilon_{h}}}\bigg \}\left(1-e^{-2\gamma\theta}\right),
\label{WQRCIT}\\
\langle WQ \rangle^{\mathrm{RM}}&=& \langle WQ \rangle^{\mathrm{RC}} - 2\Sigma^{2},
\label{WQRMIT} 
\end{eqnarray}
\end{widetext}
where $\theta=\epsilon_h\tau_h$ with $\tau_h$ being the duration of the hot bath stroke. A dependence of the moments on the initial coherence parameter $q$ is only seen in higher order $\Sigma^2$ corrections. From the expressions of these moments the according variances and the covariance of work and heat can be obtained. Instead of these bulky expressions we present the variances and covariances for the two monitoring schemes in Fig.~\ref{fig_cov}{\bf a}, {\bf b} as  functions of $\theta$. The difference between repeated measurements and repeated contacts is governed by damped oscillations stemming from the $e^{-\gamma \theta} \cos\left(2 (\theta+\varphi)\right)$-term in the deviation of the first two work moments, see Eqs.~(\ref{WRCIT}), (\ref{W2RCIT}) and Fig.~\ref{fig_cov}{\bf c}, {\bf d}. Yet, the maximal amplitude of these oscillations is rather small rendering the (co)variances for repeated measurements and repeated contacts almost identical.     

The moments of heat can be obtained by setting $\epsilon_c = 0$ and multiplying appropriate global signs (negative for odd order moments and positive for even ones) in the expressions for the moments of work. As seen from the above equations, this also implies that the average heat for two-pointer repeated contacts and repeated measurements are the same, i.e., $\langle Q\rangle = \langle Q\rangle^{\mathrm{RM}} = \langle Q\rangle^{\mathrm{RC}}$. The variances of the heat \sout{are not displayed. They} behave in a similar way as \sout{the} those of the work as presented in Fig.~\ref{fig_cov} with solely a constant difference of $3 \Sigma^2$ between repeated measurements and repeated contacts and therefore are not extra displayed. Note that the engine may starts at an arbitrary initial state specified by the parameters $d$ and $q$. Here and in the following examples, however, we will initiate the monitored sequence of $N$ cycles with the invariant state $\rho^*$ of the map $\mathcal{F}$, i.e.
\begin{equation}
\rho^* = \mathcal{F}(\rho^*)
\label{rFr}
\end{equation}
where 
\begin{equation}
\mathcal{F}(\rho) = \Phi_c( \tilde{U} \Phi_h(U \rho U^\dagger)  \tilde{U}^\dagger)
\end{equation}
describes the dynamics of a complete cycle in the absence of any monitoring.

Furthermore we note that the master equation~(\ref{Lindblad}), which governs the thermalization strokes of the engine, leads to decoupled equations of motion for the populations and coherences. Hence, the thermalization maps $\Phi_u$ resulting as the solutions of the master equation (\ref{Lindblad}) obey the conditions
\begin{equation}
\Tr [L^\pm_u \Phi_u(\mathcal{P}^k_u)] =\Tr [\mathcal{P}^k_u \Phi_u(L^\pm_u)]=\Tr [L^\pm_u \Phi_u(L^\mp_u)] =0,
\label{DD}
\end{equation}
with swap-operators $L^\pm_u = |\pm_u\rangle\langle \mp_u|$, as defined below Eq. (\ref{Lindblad}), and $\mathcal{P}^k_u =|k_u\rangle \langle k_u |$ with $k=\pm$, denoting the projection operators onto the eigenstates of the Hamiltonian $H_u$. Due to this decoupling, the work PDFs for one- and two-pointer repeated contacts become identical and we denote the average value as $\langle W\rangle^{\mathrm{RC}}$ (see Append.~\ref{Append_RedMap} for more details). In the limit $\theta \rightarrow \infty$, $\Sigma=0$, and $\rho=\rho_c$, which leads to perfect thermalization, Eqs. (\ref{WRMIT})--(\ref{WQRMIT}) match Eqs. (\ref{PT_woutput})--(\ref{PT_WQ}). For a highly non-adiabatic work stroke ($\alpha$ large) the cosine terms in the expressions for the first two moments play a significant role. Then the sign of $\cos(2(\theta+\varphi))$ determines which one of the two monitoring schemes leads to a larger work on average for a single cycle. This affects also the efficiency of the engine because the average heat is the same for either scheme.

\section{Numerical results} \label{NR}
For a working substance with a $d$-dimensional Hilbert space, the number of possible realizations of $(\vec{m}, \vec{m}^\prime)$ increases exponentially as $d^{8N}$, and so do the coefficients $D^{\vec{m},\vec{m}^\prime}$. Hence, the computational resources (CPU time and RAM) required to track the individual coefficients become inaccessible even for $d=2$ already at a moderately large number of cycles. To avoid this numerical impossibility, we device a scheme to reduce the computational complexity to $N^2$ by grouping terms according to their work values. The scheme (see Append.~\ref{Append_Compred} for details) is restricted to a two-level working substance undergoing  thermalization strokes that are governed by decoupled population and coherence dynamics as specified by Eq. (\ref{DD}). Under these conditions the method is exact and allows us to obtain the work distributions for a large number of cycles such that the asymptote can be reached.

\subsection{One cycle, perfect thermalization}\label{1cpth}
 \begin{figure}[t!]
\includegraphics[width=\columnwidth]{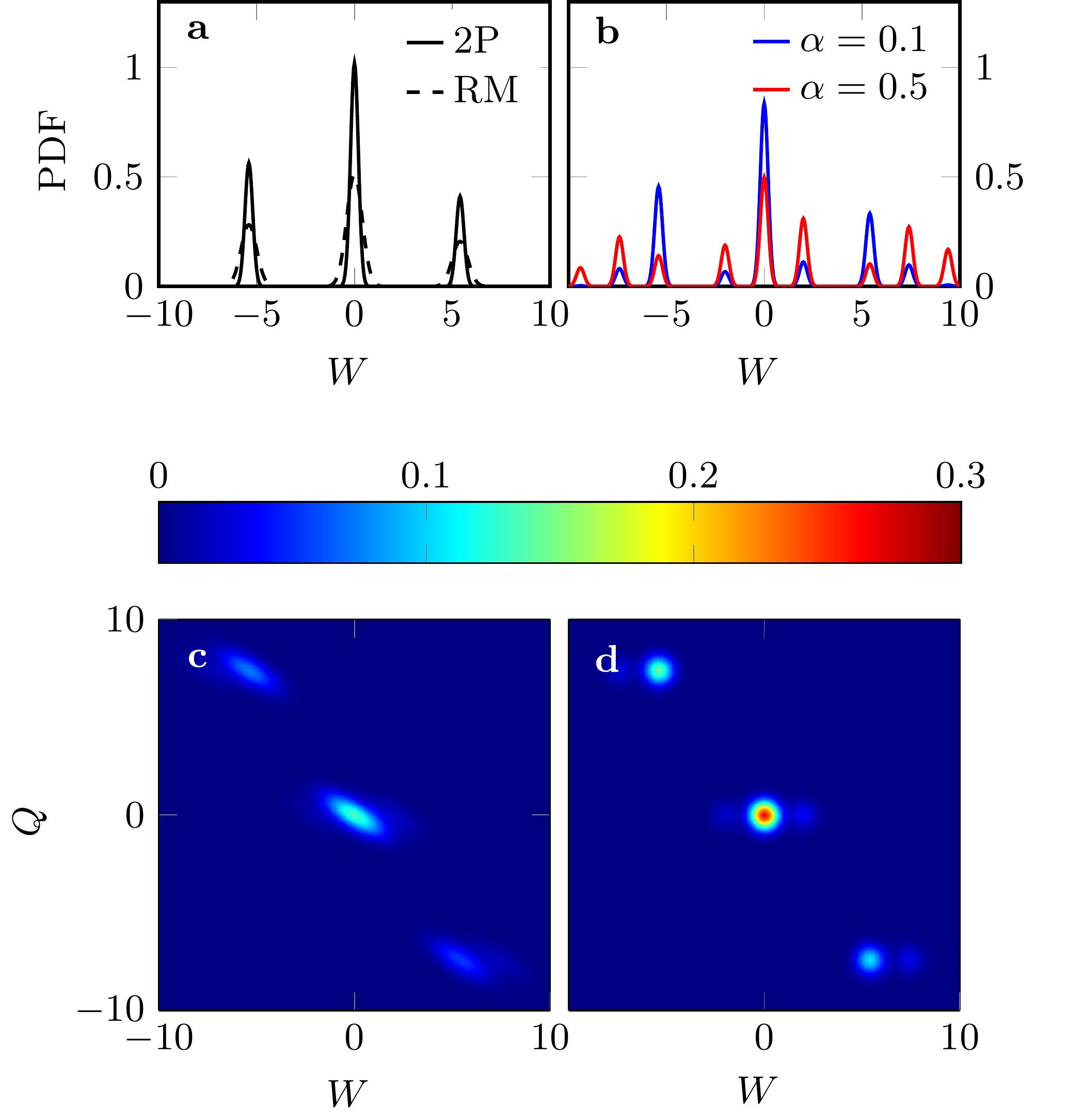}
\caption{Panels {\bf a--b} display the probability density functions (PDFs) of work for adiabatic work strokes ($\alpha =0$) and  different measurement schemes ({\bf a}) as well as  for different non-adiabatic work strokes ({\bf b}). Solid lines represent two-pointer repeated contacts PDFs, and the dashed line 
presents the PDF for repeated measurements. One-pointer results are not drawn since they are exactly the same as two-pointer scheme for $\alpha=0$ and have a negligible difference for $\alpha\neq0$. Panel {\bf c} displays the joint PDF of work and heat calculated for repeated measurements, and {\bf d} is the joint PDF calculated for two-pointer repeated contacts. The PDFs  are computed for perfect thermalization with thermal baths at inverse temperatures $\beta_c= 0.25$, $\beta_h= 0.025$, both characterized by the the same Ohmic spectal density with Lorentz-Drude cut-off frequency $\omega_{D}=0.2$, system-operator $S=\sigma_z+\sigma_x$, and damping coefficient $\gamma=0.5$. This results in density matrices with an excited-state probability $d_c =0.37759$ and $d_h = 0.45388$, and coherence $q_u = \langle +_u |\rho_u| -_u \rangle$ given by $q_h=-1.9205\times10^{-6}$ and $q_c =5.0813\times10^{-5}$ (see~Append.~\ref{Append_CPT} for details). The phase during the non-adiabatic evolution is set to $\varphi=0$ in all cases, and the joint PDF is shown for $\alpha = 0.1$. The engine is started in the generalized Gibbs state of the cold bath with parameters $d_c$ and $q_c$. All other parameters are the same as in Fig.~\ref{fig_cov}.}
\label{Fig1}
\end{figure}
Figure~\ref{Fig1} presents the joint and marginal work and heat PDFs of an engine with perfect thermalization for a single cycle ($N=1$). Panel {\bf a} displays the work PDFs for repeated measurements and repeated contacts for adiabatic work strokes, i.e., for $\alpha =0$. In this case, there is no difference between the marginal two-pointer work PDF given by Eq. (\ref{PRC2P}) and the one-pointer work PDF (\ref{PRC1P}) as explained in Append.~\ref{Append_Equiv}. In accordance with the Eq. (\ref{PRC1P}) the PDF consists of three peaks located at $W=0$ and $W=\pm 2(\epsilon_c - \epsilon_h)$ because transitions between different energy levels are impossible for adiabatic work strokes. These peaks are  broader for repeated measurements compared to repeated contacts. This feature prevails for non-adiabatic work strokes. In panel {\bf b}  we present the results for moderately and strongly non-adiabatic work strokes. In order not to clutter the plot we present only the work PDFs resulting from repeated contacts. With decreasing adiabaticity (increasing $\alpha$) six further peaks appear in accordance with Eq.~(\ref{PRC1P}) accompanied by a decrease of the weights of the three adiabatic peaks. At the same time the weights of those peaks at positive work values may increase to such an extent that the average work becomes positive. That means that the engine only consumes energy rather than to perform work. As for classical engines any non-adiabaticity is detrimental  with respect to the efficiency but must be accepted to achieve finite power. The same features as described for repeated contacts  also hold for the repeated measurement PDFs with the only difference that their peaks are broader.
         
Figures~\ref{Fig1}{\bf c} and {\bf d} present the joint  work and heat PDFs for repeated measurements and contacts, respectively. In the case of repeated measurements, each individual Gaussian contribution contains a negative covariance indicating an anti-correlation of the according work and heat contributions. In contrast, for repeated contacts the individual Gaussian contributions are  isotropic. In both cases the main contribution to the covariance of work and heat results from the linear superposition of the individual Gaussian distributions.

\subsection{$N$ cycles, imperfect thermalization}\label{IT}
\begin{figure}[t!]
\centering
\includegraphics[width=\columnwidth]{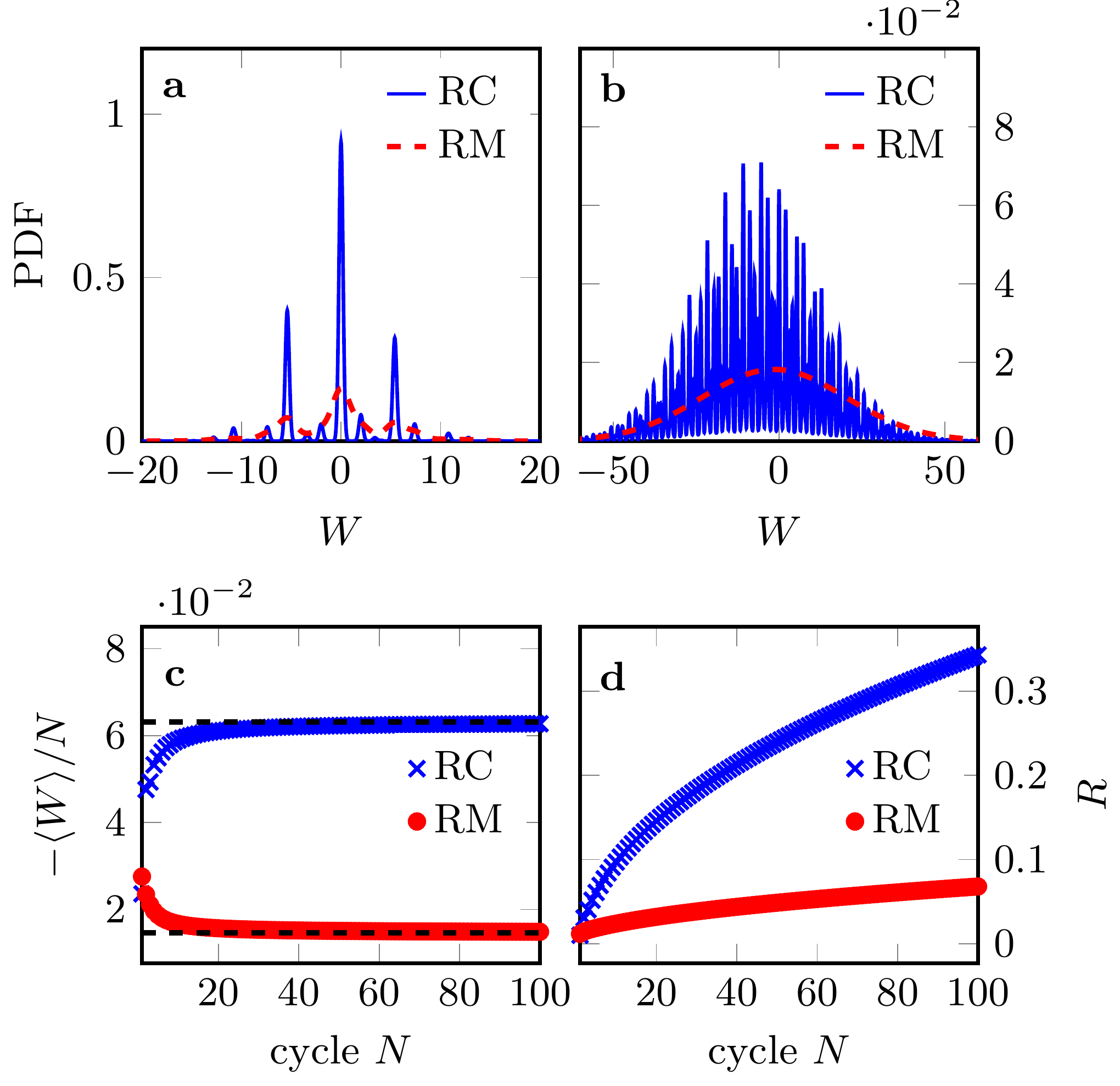}
\caption{Panel {\bf a} presents a plot of the work PDFs for repeated measurements (red) and repeated contacts (black) for $N=5$ cycles and panel {\bf b} does so for $N=100$. Two-pointer and one-pointer repeated contacts give exactly the same PDFs of work due to the dynamical decoupling between populations and coherences during the thermalization process (see text for details). While for repeated contacts separate sharp lines with $N$-independent widths remain visible also for large numbers of cycles, the according terms for repeated measurements contribute with a width $\propto \sqrt{N}$ and hence merge in a broad, practically Gaussian PDF. Panel {\bf c} and {\bf d} display the average work output per cycle and the engine's reliability, respectively. The horizontal black dashed lines in {\bf c} correspond to the asymptotic values of the average work per cycle obtained from the analytic maps described in~Append.~\ref{Append_RedMap}. The reliability scales as the square root of the cycle number $N$ for large $N$. 
The repeated measurement scheme has a detrimental impact on the engine in terms of both average work output and reliability. The work strokes are characterized by non-adiabatic parameters $\alpha=0.05$, $\varphi=0$. The energy gaps are chosen as $\epsilon_{c}=1$ and $\epsilon_{h}=3.7$. The bath parameters are $\beta_{c}=0.25$, $\beta_{h}=0.025$, the coupling constant $\gamma = 0.025$, and the thermalization times is taken such that $\theta=\epsilon_c\tau_c=\epsilon_h\tau_h=8$. All pointer states initially are Gaussian with a width of $\Sigma=0.2$. The engine is started in the according invariant state $\rho^*$ [Eq.~(\ref{rFr})] of the unobserved cycle dynamics given by the Eq.~(\ref{Fr}).}
\label{Fig2}
\end{figure}
Numerical results for adiabatic and non-adiabatic work strokes and imperfect thermalization are presented in Fig.~\ref{Fig2}. 
The panels {\bf a} and {\bf b} display the work PDFs for repeated measurements and repeated contacts for two different numbers of cycles $N$. For the smaller number of cycles, $N=5$, still several peaks are visible in the repeated measurements work PDF. Because their widths grow as $\sqrt{N}$ this multiply peaked structure disappears with increasing number of cycles to finally approach a Gaussian shape, which in the present case is virtually indistinguishable from the exact PDF for $N=30$. In contrast, the fine-structure of the repeated contact work PDF remains visible also for large numbers of cycles due to the fact that the widths of the individual Gaussian contributions are solely determined by the initial pointer state and hence  are $N$ independent.       

In Fig.~\ref{Fig2}{\bf c} the average work per cycle is displayed as it results as a function of the number of cycles both for repeated measurements (red closed circles) and repeated contacts (blue crosses). The initial state of the working substance is chosen as the asymptotic state of an unobserved engine reached after sufficiently many cycles. Only for a single cycle ($N=1$) the average work for repeated measurements is larger than that for repeated contacts. For any larger number of cycles the repeated contact scenario outperforms the repeated measurements both with respect to the average work (see panel {\bf c}) and the reliability (see panel {\bf d}). As a function of the number of cycles, the work  behaves like a biased random walk in that its first and second moments asymptotically grow in proportion to the number of cycles yielding a constant work per cycle and a reliability that increases as the square root of the number of cycles. 

For repeated measurements a map can be constructed that assigns the density matrix of the working substance after a cycle to any initial density matrix. The map possesses a unique invariant state describing the asymptotic regime of a large number of cycles. This state can be used to calculate the asymptotic average work per cycle, see~Append.~\ref{Append_RedMap}. Even though such a map of the reduced working substance density matrix does not exist for repeated contacts, in the special cases of the Lindblad equation (\ref{Lindblad}) that dynamically decouples populations and coherences, the asymptotic value of the average work can be determined from the asymptotic density matrix in the absence of any contacts or measurements, see~Append.~\ref{Append_RedMap}. This unobserved asymptotic density matrix of the working substance can again be  obtained as the invariant state of a  quantum map in the absence of contacts as considered in Ref.~\cite{Kosloff17}. The resulting average works are presented in panel {\bf c} as dashed black lines.        

\subsection{Power of a Landau-Zener engine}\label{Ex}
So far we have characterized the two work strokes by  the Hamiltonians of the expanded and the compressed working substances [see Eqs. (\ref{Shevchenko10}) and (\ref{phi})] and unitary operators specifying the  changes of its state, however, without specifying a particular forcing protocol. As a consequence, the duration of the work strokes is not specified. In general, determining the unitary time evolution operator for a particular forcing protocol of given duration presents a nontrivial task~\cite{Barnes12, Bandyopadhyay21}. In order to also have full control over the duration of an engine cycle we consider a forcing protocol with a linear driving as specified in Eq. (\ref{Shevchenko10}). The unitary time evolution for a work stroke of the duration $T_1/2$ is then determined by the parameters as given in the Eq.~(\ref{phi}). The duration of the thermalization strokes, $\tau_h$ and $\tau_c$ are  assumed to be fixed by a the single parameter $\theta= \epsilon_c \tau_c =  \epsilon_h \tau_h$, resulting in the total thermalization time 
\begin{equation}
T_2 = \tau_c +\tau_h =\theta(1/\epsilon_h + 1/\epsilon_c)
\label{T2}
\end{equation} 
Therefore, a single cycle takes the time $T_1+T_2$.  The power of the engine is given by
\begin{equation}
P = -\frac{\langle W\rangle}{T_{1}+T_{2}}.
\end{equation}
where $\langle W \rangle$ denotes the average work performed on the working substance per cycle.
\begin{figure}[t!]
\centering
\includegraphics[width=\columnwidth]{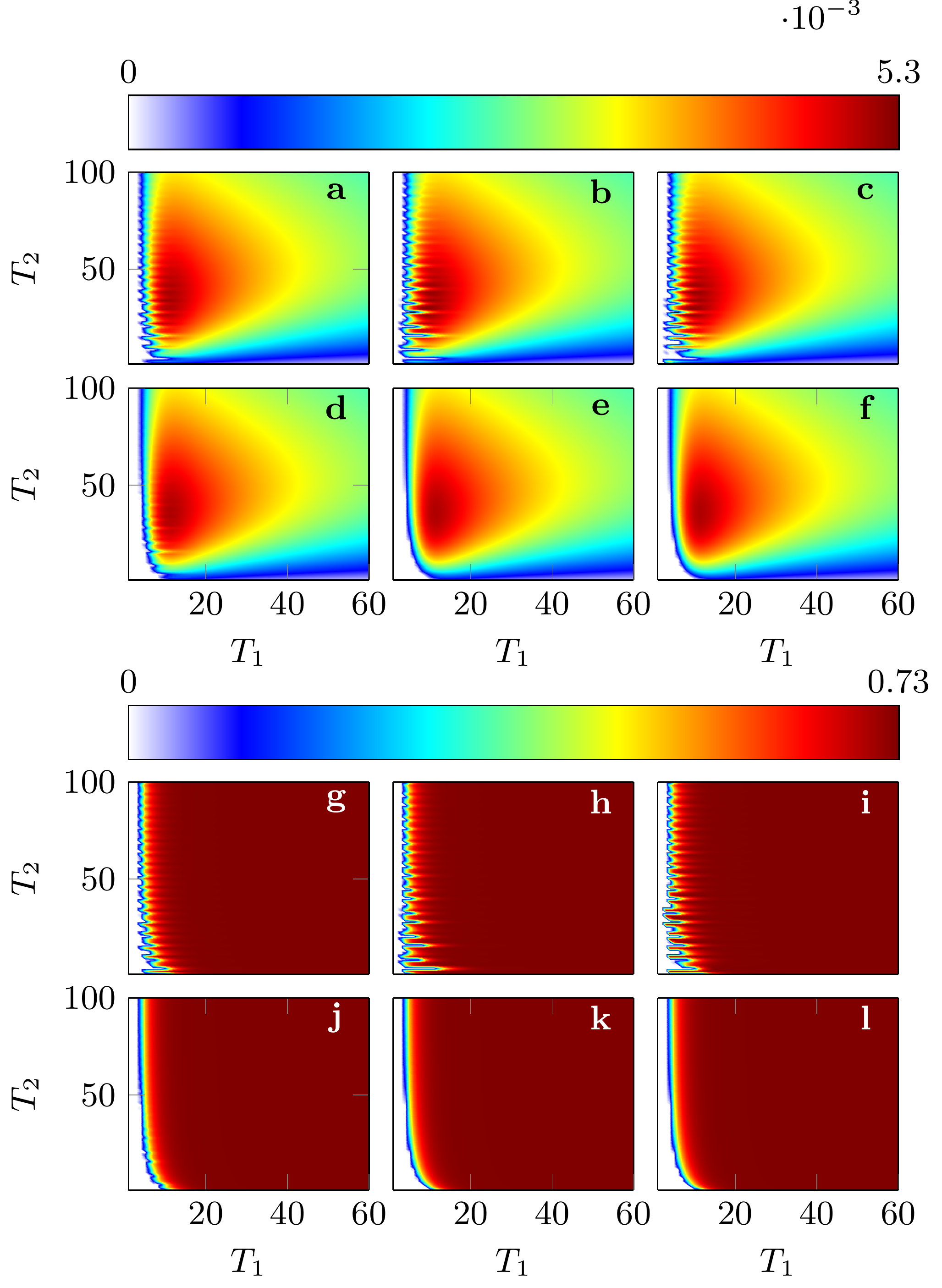}
\caption{Power per cycle ({\bf a}-{\bf f}) and average efficiency ({\bf g}-{\bf l}) as a function of work and heat stroke durations $T_1$ and $T_2$ with cycle number $N=1$ ({\bf a}, {\bf d}, {\bf g}, and {\bf j}), $N=20$ ({\bf b}, {\bf e}, {\bf h}, and {\bf k}), and $N=\infty$ ({\bf c}, {\bf f}, {\bf i}, and {\bf l}; see~Append.~\ref{Append_RedMap}). Panels {\bf a}-{\bf c} and {\bf g}-{\bf i} are for the engine measured via repeated contacts whereas the other panels are for repeated measurements. The white region in all panels correspond to the dud regime of the engine. 
The extended dark red regions in {\bf g}--{\bf l} indicate regimes with efficiencies close to the ideal Otto engine value $1-\epsilon_{c}/\epsilon_{h} \approx 0.73$. Fixed parameters are $\epsilon_{c} = 1$, $\epsilon_{h} = 3.7$, $\theta=\epsilon_c\tau_c=\epsilon_h\tau_h=8$, $\beta_{c} = 0.25$, $\beta_{h}=0.025$, $\gamma = 0.025$, and $\Sigma = 0.2$. The engine is always started in the invariant state $\rho^*$ [Eq.~(\ref{rFr})] of the according unobserved cycle dynamics given by the Eq.~(\ref{Fr}).   
}
\label{Fig3}
\end{figure}

In Fig.~\ref{Fig3} the power and the efficiency for repeated contacts and repeated measurements as well as for different numbers of cycles are presented as functions of the duration of the work and thermalization strokes. The variance of the pointer states is chosen relatively small such that coherences that may build up during the non-adiabatic work strokes and also partially during an imperfect thermalization, are almost completely suppressed by repeated measurements while they can partly survive in the presence of repeated contacts. Because the generation of coherences is further suppressed in the limit of adiabatic work processes as well as for perfect thermalization, one finds substantial differences between repeated measurements and repeated contacts for small values of $T_1$ and small to intermediate values of $T_2$. As already observed in Sec.~\ref{1cpth}, the total average work performed on the working substance in a single cycle becomes positive for perfect thermalization and work strokes with a sufficiently large parameter $\alpha$ and hence for a sufficiently short duration of the work strokes. This feature of a dud engine with formally negative power and efficiency persists for imperfect thermalization and also for any numbers of cycles at sufficiently small values of the work stroke duration $T_1$ indicated as  white regions in Fig.~\ref{Fig3}. The border between the regions of a properly working and a faulty engine displays well pronounced oscillations as a function of $T_2$ for any number of cycles in the case of repeated contacts. The fact that these oscillations are visible in the case of repeated measurements only for small number of cycles indicates that they are caused by coherences which are suppressed by repeated measurements.
\begin{figure}[!t]
\centering
\includegraphics[width=\columnwidth]{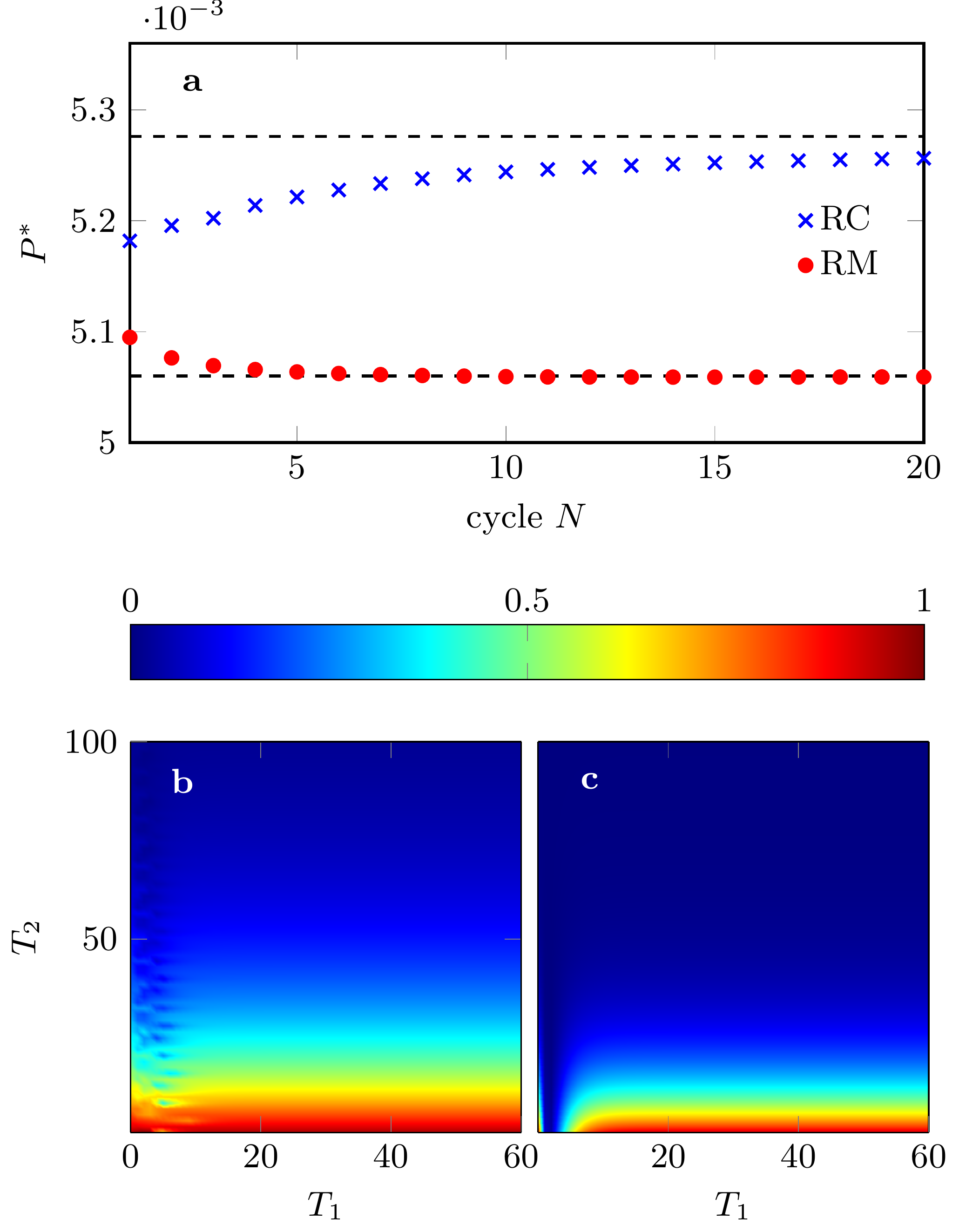}
\caption{Maximum power $P^{*}$, with respect to durations $T_1$ and $T_2$, of the engine as a function of cycle number $N$ for both repeated contacts and repeated measurements, blue and red symbols, respectively (panel {\bf a}). The dashed lines represent the asymptotic values of maximum power of the engines in their periodic state asymptotically reached after many cycles, see Append.~\ref{Append_RedMap}. The  deviations of the engine from this limiting behavior decay exponentially as the cycle number increases.  Panel {\bf b} represents the slowest decaying factor $\Lambda_{2}$ (see~\ref{Append_RedMap} last paragraph for definition) in engines monitored by repeated contacts, while {\bf c} is the corresponding plot of engines being repeatedly measured. The initial condition for the engine in panel {\bf a} is again given by the according invariant state $\rho^*$ as in the previous figures. The other parameters $\epsilon_{c} = 1$, $\epsilon_{h} = 3.7$, $\theta=\epsilon_c\tau_c=\epsilon_h\tau_h=8$, $\beta_{c} = 0.25$, $\beta_{h}=0.025$, and $\gamma = 0.025$ are the same as in Fig.~\ref{Fig3}}
\label{Fig5}
\end{figure}

The dependence of the maximal power $P^* = \max_{T_1,T_2} P$  on the number of cycles for repeated measurements and contacts is displayed in Fig.~\ref{Fig5}{\bf a}.  This maximum is chosen from the set of all engines with variable stroke times $T_1$ and $T_2$ and with all other parameters fixed.  In both cases, the power approaches an asymptotic value at large numbers of cycles. While the asymptotic limit is reached from above already after a few cycles for repeated measurements, it is approached considerably slower from below for repeated contacts. The asymptotic values are indicated as broken horizontal lines. In both cases, these values  can be determined from the invariant maps that propagate the density matrix cycle by cycle. For the repeated contact scenario, the map describes the dynamics in the absence of any monitoring, while for the repeated measurement scenario it describes the single-cycle engine dynamics  in the presence of  non-selective energy measurements at the times indicated in Fig.~\ref{Fig0}. For a more detailed description we refer to~Append.~\ref{Append_RedMap}. 

For the considered parameter values of the engine, the position where the maximum power is assumed in the $T_1$-$T_2$ plane turns out to be almost independent of the number of cycles. Therefore, the convergence speed of the  map as it is iterated, mainly determines the convergence of the power with increasing number of cycles as it can be seen in Fig.~\ref{Fig5}{\bf a}. For sufficiently many cycles the convergence speed is ruled by the second largest eigenvalue of the respective map, while its largest eigenvalue is given by $\Lambda_1=1$. It is non-degenerate for all engines with a positive thermalization time $T_2>0$. The dependence of the second largest eigenvalue $\Lambda_2$  as a function of $T_1$ and $T_2$ is exemplified in the panels   Fig.~\ref{Fig5}{\bf b},{\bf c}. In both scenarios the second eigenvalue converges for large  thermalization times to zero, however much faster for repeated measurements than for repeated contacts in agreement with the faster convergence of the maximal power towards its asymptotic limit, i.e., $\Lambda^{RM}_2 < \Lambda^{RC}_2$. Outside the regime of short work strokes,  the second eigenvalue is almost independent of $T_1$ as it is primarily determined by the duration and strength of the thermalization strokes. 

\section{Conclusions}\label{Conclusions}

In this paper, we compared two different monitoring schemes, both designed to record the statistics of work performed on and heat supplied to a quantum Otto engine over a specified number of engine cycles. Based on these statistics, we investigated the performance of such engines running subject to different modes of operation, such as different work strokes ranging from adiabatic to sudden compression and expansion of the working substance, and different types of heating and cooling strokes. The method was formulated for general working substances and exemplified for a two-level system acting as working substance. 

Apart from the joint and marginal work and heat PDFs, we studied  performance measures such as efficiency, power and reliability that are all determined by the first two moments of work and heat. For an engine that performs $N$ cycles, the two monitoring schemes are based on contacts of the working substance with one, two, or $4N$ pointers, depending on the particular scheme. These engine-pointer contacts are supposed to be so short that the dynamics of the working substance is frozen during these moments but effective enough to induce a change of the pointer state depending on the actual energy of the system. 

For the repeated measurement scheme each contact is immediately followed by a projective measurement of the pointer state whose value corresponds to the energy of the working substance at the instant of the contact, hence representing a generalized energy measurement according to the von Neumann scheme. The precision of the measurement can be controlled by the position variance $\Sigma^2$ of the initial pointer states which we supposed to be Gaussian and identical for all employed pointers. From the joint statistics of all measured energy values, the joint and marginal PDFs of work and heat can be found. These PDFs result as linear superpositions of Gaussian contributions with centers at work and heat values given by proper sums and differences of the $4N$ energy values but also fictitious values resulting from a double set of energies. The latter though are exponentially suppressed provided the variance $\Sigma^2$ is sufficiently small. All contributing Gaussians are of the same shape with a variance growing proportionally to the number of cycles leading to a smeared-out, almost Gaussian PDF for sufficiently large numbers of cycles.

For repeated contacts there are just two pointers one being associated to work and the other one to heat. These pointers are ``kept on'' during the whole monitored period of cycles. At each contact the pointer state is modified according to the present energetic state of the working substance, left unchanged during the strokes and read out by  projective position measurements only after the completion of $N$ cycles. The energetic states at the subsequent contacts are inscribed  with alternating signs in the work pointer. For the heat pointer only those contacts before and after a heat stroke are registered with appropriate signs. It is interesting to note that the marginal distributions of work obtained from the joint work and heat PDF in general differ from the work PDF of a repeated contact scheme with only the work pointer but without the heat pointer. The same holds for the heat PDFs obtained either as the marginal of the joint work and heat PDF or as the outcome of a single heat pointer repeated contact scheme. This aspect of quantum contextuality is absent in the repeated measurement scheme. It indicates a higher sensitivity of the repeated contact scheme to the quantumness of a process as compared to the repeated measurement scheme.           

The higher sensitivity of the repeated contact scheme is closely related to the lesser back-action of the sole contacts as compared to the individual read-out of the pointer-state position for repeated measurements leading to different suppression factors making long-lived coherences possible. In the example of an Otto engine with a two-level system working substance and imperfect thermalization strokes this manifests itself in an average work per cycle that first increases with the number of cycles and finally saturates at a constant value for repeated contacts. In contrast, for repeated measurements the average work per cycle decreases towards an asymptotic value that is smaller than that for repeated contacts. Accordingly, the power but also the efficiency as well as the reliability of the work, defined as the ratio of the average work and its root-mean-square deviation,  are larger for repeated contacts than for repeated measurements provided that the engine protocol allows coherences being still present after a cycle. Due to the less strong back-action imposed by repeated contacts in comparison to repeated measurements these coherences may subsist leading to a better performance of a repeated-contacts-monitored engine. We expect that this behaviour is not restricted to the special case of a two-level working substance but that it prevails in general. 

In a recent work, Watanabe et.~al. \cite{Watanabe17} studied the amount of energy a quantum Otto engine may deposit into an external system over several engine cycles. The comparison of the amount of out-coupled energy into a ``flywheel'' with the energy produced by the free engine according to a repeated contact diagnostic scheme may provide further insight into the effectiveness of charging and discharging strategies of quantum batteries.    

\section*{Acknowledgements}
This research was supported by the Institute for Basic Science in Korea (IBS-R024-Y2).
\appendix

\section{Generalized Gibbs state}\label{Append_CPT}
The generalized Gibbs state can be obtained up to second order in system-bath interaction using canonical perturbation theory \cite{Thingna12}. In this Appendix we omit the subscripts $c$ or $h$ used in the main text for the cold and hot reservoir for notational simplicity. The generalized Gibbs state 
\begin{equation}
\rho = \frac{\Tr_{B}\left[e^{-\beta H_{\mathrm{tot}}}\right]}{\Tr\left[e^{-\beta H_{\mathrm{tot}}}\right]},\label{ggt_general}
\end{equation}
can be interpreted as the reduced density matrix of an extended system governed by the total Hamiltonian $H_{\mathrm{tot}}=H_S + H_B + V$ in a canonical state at the inverse temperature $\beta$. Here, $H_S$ and $H_B$, denote the Hamiltonians of the proper system and the bath, respectively and $V$ denotes the interaction. We assume that the bath is composed of independent Harmonic oscillators of frequency $\omega_n$ and the system-bath interaction takes the form $V=-S\otimes \sum_n c_n x_n$, with $S$ being a system operator and $x_n$ being the position operator of the $n$th mode of the bath coupling to the system via strength $c_n$. Performing a perturbative expansion upto second-order in $V$ we obtain,
\begin{equation}
\rho = \frac{e^{-\beta H_{S}}}{Z_{S}} + \frac{D}{Z_{S}} - \frac{e^{-\beta H_{S}}\Tr[D]}{Z_{S}^{2}}.
\end{equation}
Above $Z_S = \Tr[\exp[-\beta H_S]]$ is the system partition function of the bare system and the operator $D$ encodes information about finite system-bath coupling, i.e.,
\begin{eqnarray}
D &=& e^{-\beta H_{S}}\bigg[\int_{0}^{\beta}\,d\beta_{1}\int_{0}^{\beta_{1}}\,d\beta_{2}\,\tilde{S}_{\beta_{1}}\tilde{S}_{\beta_{2}}C(-i(\beta_{1}-\beta_{2}))\bigg].\nonumber
\end{eqnarray}
The operator $\tilde{S}_{\mathrm{x}} = e^{\mathrm{x} H_S} S e^{-\mathrm{x} H_S}$ is the imaginary time free evolution of the system operator and the force-force correlator $C(\tau) = \sum_{n,n'}c_n c_{n'}\Tr[e^{-\beta H_B}\tilde{x}_n(\tau) x_{n'}]/\Tr[e^{-\beta H_B}]$ with $\tilde{x}_n(\tau) = e^{i\tau H_B} x_n e^{-i\tau H_B}$. The force-force correlator can be expressed in terms of the spectral density of the bath, 
\begin{equation}
J(\omega) = \pi\sum_{n}\frac{c_{n}^{2}}{2m_{n}\omega_{n}}\delta(\omega-\omega_{n}),
\end{equation}
as
\begin{eqnarray}
C(\tau) &=& \int_{0}^{\infty}\,d\omega\frac{J(\omega)}{\omega}\left[\coth\left(\frac{\beta\omega}{2}\right)\cos(\omega\tau) - i\sin(\omega\tau)\right].\nonumber
\end{eqnarray}
For our perfectly thermalizing two-level Otto engine discussed in Sec.~\ref{PT} we use the system operator $S = \sigma_{z}+\sigma_{x}$ such that the density matrix
\begin{eqnarray}
\rho & = & d |+\rangle\langle +| + (1-d) |-\rangle\langle -| + [q |+\rangle \langle - | + \mathrm{h.c.}],
\label{rhoth}
\end{eqnarray}
with
\begin{eqnarray}
d &=& \frac{e^{-\beta \epsilon}}{Z_S} \left(1+ 4\int_{0}^{\beta}\,d\lambda \frac{(\beta - \lambda)  \sinh(2\epsilon\lambda)}{(1+e^{-2\beta\epsilon})}C(-i\lambda)\right), \nonumber \\
q &=&\frac{e^{-\beta \epsilon}}{Z_S}\int_{0}^{\beta}\,d\lambda \frac{e^{2\beta\epsilon}(e^{-2\epsilon\lambda}-1)+(e^{2\epsilon\lambda}-1)}{\epsilon}C(-i\lambda),\nonumber \\
\label{d}
\end{eqnarray}
with $Z_S = e^{\beta\epsilon} + e^{-\beta\epsilon}$. Throughout this work we have used an Ohmic spectral density with Lorentz-Drude cut-off as $J(\omega) = \gamma\omega/(1+\omega^2/\omega_D^2)$ for all our numerical simulations.

\section{Reduced maps for measured engines}\label{Append_RedMap}
In this section, we present the methods to determine the map for the average work performed in the $(N+1)$th cycle based on the knowledge of the average work performed during $N$ cycles. In case of repeated measurements, the map fully captures the distribution of work performed at any $(N+1)$th cycle, whereas for repeated contacts we introduce a \emph{fictitious} work distribution whose average is equivalent to the asymptotic average work per cycle. The method is explained for work performed in the last cycle but can likewise be applied for the heat absorbed in the last cycle.       

\subsection{Repeated measurements}
\label{Append_RedMapRM}
First, we consider the case of repeated measurements. As described in Sec.~\ref{dmI}, in this case after every stroke a new Gaussian pointer is adopted, brought into contact with the engine, and measured. We begin with the non-normalised density matrix for $(N+1)$ cycles, using Eq.~(\ref{RM_tot}),
\begin{widetext}
\begin{equation}
\phi^{\mathrm{RM}}_{\vec{x}_{N+1}}(\rho) = \sum_{\vec{m}_{N+1},\vec{m}^{\prime}_{N+1}}D^{\vec{m}_{N+1},\vec{m}^{\prime}_{N+1}}(\rho)\prod_{k=1}^{4(N+1)}\sigma \left(x_{k}-e_{k}^{m_{k}},x_{k}-e_{k}^{m_{k}^{\prime}}\right).
\label{eq:B1}
\end{equation}
\end{widetext}
Utilising the relation in Eq.~(\ref{Dmap}), splitting the product above into one for $N$ cycles and the other for $(N+1)$th cycle, and using the non-normalized density matrix for $N$ cycles [Eq.~(\ref{RM_tot})] we can rewrite the above equation as,
\begin{eqnarray}
\phi_{\vec{x}_{N+1}}^{\mathrm{RM}}(\rho) &=& \sum_{\mathbf{m},\mathbf{m}^{\prime}}S^{\mathbf{m},\mathbf{m}^{\prime}}\left(\phi_{\vec{x}_{N}}^{\mathrm{RM}}(\rho)\right) \nonumber \\
&&\times \prod_{k=4N+1}^{4(N+1)}\sigma\left(x_k-e_k^{m_k},x_k-e_k^{m_k^\prime}\right).
\label{eq:C2}
\end{eqnarray}
The work performed in the $(N+1)$th cycle is given by
\begin{equation}
\label{eq:WN}
    W^{(N+1)} = \sum_{k=4N+1}^{4(N+1)}(-1)^{k}e_{k},
\end{equation} 
and its corresponding PDF is
\begin{widetext}
\begin{eqnarray}
    P^{\mathrm{RM}}(W^{(N+1)}) &=& \int d^{4(N+1)}x\,\delta\left(W^{(N+1)}-\sum_{k=4N+1}^{4(N+1)}(-1)^{k}x_{k}\right)p(\vec{x}_{N+1}),
\end{eqnarray}
\end{widetext}
with $p(\vec{x}_{N+1}) = \Tr[\phi_{\vec{x}_{N+1}}^{\mathrm{RM}}(\rho)]$ according to Eq.~(\ref{pxphi}). Noting that the integration variables within the delta function depend only on the $(N+1)$th cycle pointer positions 
$x_k$, the integral can be split as,
\begin{widetext}
\begin{equation}    
   P^{\mathrm{RM}}(W^{(N+1)})= \int d\mathbf{x}\,\delta\left(W^{(N+1)}-\sum_{k=4N+1}^{4(N+1)}(-1)^{k}x_{k}\right)\int d^{4N}x \,p(\vec{x}_{N+1}),
   \label{NcycWPDFRM}
\end{equation}
\end{widetext}
with $\mathbf{x}$ defined by $\vec{x}_{N+1} = \vec{x}_{N}\oplus\mathbf{x}$. All information about the previous $N$ cycles is contained in the rightmost integral above. Hence, using Eq.~(\ref{eq:C2}) this integral can be expressed as,
\begin{eqnarray}
\int d^{4N}x \, p(\vec{x}_{N+1}) &=& \sum_{\mathbf{m},\mathbf{m}^{\prime}}\Tr\left[S^{\mathbf{m},\mathbf{m}^\prime}\left(\int d^{4N}x\, \phi_{\vec{x}_{N}}^{\mathrm{RM}}(\rho)\right)\right]\nonumber\\
    && \times\prod_{k=4N+1}^{4(N+1)}\sigma\left(x_k-e_k^{m_k},x_k-e_k^{m_k^\prime}\right).\nonumber
\end{eqnarray}
The simplification on the r.h.s. implies that as long as we know the map $S^{\mathbf{m},\mathbf{m}^\prime}$ and the state it acts on, i.e., $\int d^{4N}x\, \phi_{\vec{x}_{N}}^{\mathrm{RM}}(\rho)$, we can evaluate the work distribution $P^{\mathrm{RM}}(W^{(N+1)})$. Thus, the task reduces to finding the map $F^{\mathrm{RM}}$, such that,
\begin{equation}
    \int d^{4N}x\,\phi_{\vec{x}_N}^{\mathrm{RM}}(\rho) = F^{\mathrm{RM}}\left(\int d^{4N-4}x\,\phi_{\vec{x}_{N-1}}^{\mathrm{RM}}(\rho)\right).
    \label{eq:B7}
\end{equation}
Using Eq.~(\ref{eq:C2}) with $N\rightarrow N-1$ and integrating both sides we can easily obtain the map,
\begin{equation}
    F^{\mathrm{RM}}(\varrho) = \sum_{\mathbf{\tilde{m}},\mathbf{\tilde{m}}^{\prime}}S^{\mathbf{\tilde{m}},\mathbf{\tilde{m}}^{\prime}}(\varrho)\prod_{k=4N-3}^{4N}e^{-\frac{1}{8\Sigma^{2}}\left(e_k^{m_k}-e_k^{m_k^\prime}\right)^{2}},\label{DRM}
\end{equation}
where $\varrho = \int d^{4N-4}x\,\phi_{\vec{x}_{N-1}}^{\mathrm{RM}}(\rho)$, ${\bf \tilde{m}^{(\prime)}}$ are defined by the sequence $\vec{m}^{(\prime)}_N = \vec{m}^{(\prime)}_{N-1} \oplus {\bf \tilde{m}^{(\prime)}}$. Note the difference between ${\bf \tilde{m}^{(\prime)}}$ and ${\bf m^{(\prime)}}$ is that $N$ is replaced by $(N+1)$ in the concatenation. The map $F^{\mathrm{RM}}$ is \emph{Markovian} since it gives us the state at the $N$th cycle given the state at $N-1$ [see Eq.~(\ref{eq:B7})]. Overall, in case of repeated measurements since the map $F^{\mathrm{RM}}$ is known it is straightforward to obtain the work PDF for the work performed in the $(N+1)$th cycle using Eqs.~(\ref{NcycWPDFRM})--(\ref{DRM}). 

\subsection{Repeated contacts}
\label{Append_RedMapRC}
In case of repeated contacts, one cannot assign a value of work to each individual cycle as we did with $W^{(j)}$ for repeated measurements by using the classically stored information about the outcomes of the sequence of energy measurements for $N$ cycles. To circumvent this problem  we restrict our consideration to engines with a working substance consisting of a two-level system and moreover being subject to  thermalization strokes obeying the decoupled population and coherences dynamics specified by the Eq.~(\ref{DD}). In this case we are able to introduce a fictitious work PDF that can be computed recursively and that determines the average work per cycle in the limit of large cycle numbers when the engine has reached a steady mode of operation.   

In order to proceed, we begin with the work PDFs defined in Eqs.~(\ref{PRC2P}) and (\ref{PRC1P}),
\begin{widetext}
\begin{eqnarray}
 P^{\mathrm{RC,2P}}(W)&=& \sum_{\vec{m},\vec{m}^\prime} D^{\vec{m},\vec{m}^\prime}  e^{-\frac{1}{8 \Sigma^2} (\mathcal{W}^{\vec{m}} - \mathcal{W}^{\vec{m}\prime})^2} e^{-\frac{1}{8 \Sigma^2} (\mathcal{Q}^{\vec{m}} - \mathcal{Q}^{\vec{m}\prime})^2} g_{\Sigma^2}(W-\mathcal{W}^{\vec{m},\vec{m}^\prime}), \nonumber \\
 P^{\mathrm{RC,1P}}(W)&=& 
 \sum_{\vec{m},\vec{m}^\prime} D^{\vec{m},\vec{m}^\prime}  e^{-\frac{1}{8 \Sigma^2} (\mathcal{W}^{\vec{m}} - \mathcal{W}^{\vec{m}\prime})^2}  g_{\Sigma^2}(W-\mathcal{W}^{\vec{m},\vec{m}^\prime}).
\end{eqnarray}
\end{widetext}
In case of dynamical decoupling between populations and coherences for a two-level working substance, [Eq.~(\ref{DD})], only the terms with
\begin{equation}
    e_{2i}^{m_{2i}}-e_{2i}^{m_{2i}^{\prime}} = e_{2i+1}^{m_{2i+1}}-e_{2i+1}^{m_{2i+1}^{\prime}}\label{RC_simplifying_DD}
\end{equation}
$\forall i = 1,\cdots,2N-1$ contribute to the work distributions (see~Append.~\ref{Append_DDsimp} for details). This simplifies the exponents [see Eq.~(\ref{WRC}) and (\ref{QRC})] appearing in the work distributions, namely,
\begin{eqnarray}
    \mathcal{W}^{\vec{m}}-\mathcal{W}^{\vec{m}^{\prime}} &=& -e_{1}^{m_1}+e_{1}^{m_1^{\prime}} +e_{4N}^{m_{4N}}-e_{4N}^{m_{4N}^{\prime}},\label{RC_simplifying} \\
    \mathcal{Q}^{\vec{m}} - \mathcal{Q}^{\vec{m}\prime} &=& 0.
\end{eqnarray}
Thus, for the models considered in Secs.~\ref{IT}--\ref{Ex} the one- and two-pointer work PDFs always coincide. Moreover, using the above simplifications one can easily identify the non-normalized reduced density matrix for the working substance as,
\begin{eqnarray}
    \phi^{\mathrm{RC}}_{W}(\rho) &=& \sum_{\vec{m},\vec{m}^{\prime}}D^{\vec{m},\vec{m}^{\prime}}(\rho)\,e^{-\frac{1}{8\Sigma^{2}}\left(e_{1}^{m_1}-e_{1}^{m_1^{\prime}}\right)^{2}}\nonumber \\
    && \times g_{\Sigma^{2}}\left(W-\mathcal{W}^{\vec{m},\vec{m}^{\prime}}\right)\nonumber\\
    &=& \sum_{\vec{m},\vec{m}^{\prime}}D^{\vec{m},\vec{m}^{\prime}}(\tilde{\rho})g_{\Sigma^{2}}\left(W-\mathcal{W}^{\vec{m},\vec{m}^{\prime}}\right).
    \label{RC_W}
\end{eqnarray}
Here we also omitted the only remaining contribution $e_{4N}^{m_{4N}}-e_{4N}^{m_{4N}^{\prime}}$ to the exponential factor $e^{-(\mathcal{W}^{\vec{m}}-\mathcal{W}^{\vec{m}^{\prime}})^2/(8 \Sigma^2)}$ because the terms $D^{\vec{m},\vec{m}^{\prime}}$ with  $e_{4N}^{m_{4N}}\neq e_{4N}^{m_{4N}^{\prime}}$ vanish in the final trace operation applied to obtain the work PDFs from the  non-normalized reduced density matrices. In addition, we made use of the fact that the innermost part of the telescopic action of the map $D^{\vec{m},\vec{m}^\prime}$ on $\rho$ is given by  $\mathcal{P}^{m_1}_1 \rho \mathcal{P}^{m^\prime_1}_1$, see Eq.~(\ref{Dmap}). Splitting the density matrix $\rho$ into its diagonal part $\rho_d = \sum_{m=1,2} \mathcal{P}^m_1 \rho \mathcal{P}^m_1$ and the remaining coherence contribution $\rho_c = \rho - \rho_d$ one obtains $ \exp\left[-(e_{1}^{m_1}-e_{1}^{m_1^{\prime}})^{2}/(8\Sigma^{2})\right] \mathcal{P}^m_1 \rho \mathcal{P}^{m^\prime}_1=\mathcal{P}^m_1 \tilde{\rho} \mathcal{P}^{m^\prime}_1$ where
\begin{equation}
\tilde{\rho} \equiv \rho_d + e^{-\frac{\epsilon^2_c}{2 \Sigma^2}} \rho_s.
\label{rhods}
\end{equation}
Note that the map $\rho \to \tilde{\rho}$ preserves  the trace and the positivity.

Next we introduce the fictitious work distribution
\begin{eqnarray}
    \tilde{P}^{\mathrm{RC}}(W^{(N+1)}) &=& \int d\mathbf{x}\,\delta\left(W^{(N+1)}-\sum_{k=4N+1}^{4(N+1)}(-1)^{k}x_{k}\right) \nonumber \\
    && \times \int d^{4N}x\,\tilde{p}(\vec{x}_{N+1}),\label{PRCWN}
\end{eqnarray}
which exactly resembles Eq.~(\ref{NcycWPDFRM}) with,
\begin{eqnarray}
\tilde{p}(\vec{x}_{N}) &=& \sum_{\vec{m}_N,\vec{m}^{\prime}_N} \Tr\left[D^{\vec{m}_N,\vec{m}^{\prime}_N}(\tilde{\rho})\right] \nonumber \\
&&\times \prod_{k=1}^{4N} g_{\Sigma^2} \left (x_k - \frac{1}{2} \left (e^{m_k}_k + e^{m_k^\prime}_k\right) \right ).\label{pxN}
\end{eqnarray}
The average work from the fictitious distribution,
\begin{equation}
    \langle{W}^{(N+1)}\rangle = \int dW^{(N+1)}\,W^{(N+1)}\tilde{P}^{\mathrm{RC}}(W^{(N+1)}),
\end{equation}
represents the difference of the average work performed by the engine in the $(N+1)$th and $N$th cycle, i.e., $\langle{W}^{(N+1)}\rangle = \langle W\rangle_{N+1} - \langle W\rangle_{N}$ ( this relation follows from the definitions of the individual quantities as explained below). As the monitored engine reaches its asymptotic state, $\langle W^{(\infty)}\rangle$ represents the average work performed by the engine per cycle, similarly to the repeated measurements case, because the monitoring process on average has the same effect on the engine state in the $N$th and $(N+1)$th cycle after sufficiently many cycles.

The average work performed in $(N+1)$ and in $N$ cycles can be written as
\begin{widetext}
\begin{eqnarray*}
    \langle W\rangle_{N+1} &=& \int dW\,W\sum_{\vec{m}_{N+1},\vec{m}_{N+1}^\prime}\Tr\left[D^{\vec{m}_{N+1},\vec{m}_{N+1}^\prime}(\tilde{\rho})\right]g_{\Sigma^{2}}(W-\mathcal{W}^{\vec{m}_{N+1},\vec{m}_{N+1}^\prime})\\
    &=& \int dW\,W\sum_{\mathbf{m},\mathbf{m}^\prime}\Tr\left[S^{\mathbf{m},\mathbf{m}^\prime}(\sum_{\vec{m}_{N},\vec{m}_{N}^\prime}D^{\vec{m}_{N},\vec{m}_{N}^\prime}(\tilde{\rho})])\right]g_{\Sigma^{2}}(W-\mathcal{W}^{\vec{m}_{N},\vec{m}_{N}^\prime}-\mathcal{W}^{\mathbf{m},\mathbf{m}^\prime}),\\
    \langle W\rangle_{N} &=&  \int dW\,W\sum_{\mathbf{m},\mathbf{m}^\prime}\Tr\left[S^{\mathbf{m},\mathbf{m}^\prime}(\sum_{\vec{m}_{N},\vec{m}_{N}^\prime}D^{\vec{m}_{N},\vec{m}_{N}^\prime}(\tilde{\rho})])\right]g_{\Sigma^{2}}(W-\mathcal{W}^{\vec{m}_{N},\vec{m}_{N}^\prime}).
\end{eqnarray*}
\end{widetext}
The third line holds because the non-selective map $F^{\mathrm{RC}}$ that is defined as
\begin{equation}
F^{\mathrm{RC}}(\varrho)= \sum_{\mathbf{\tilde{m}},\mathbf{\tilde{m}}^\prime}S^{\mathbf{\tilde{m}},\mathbf{\tilde{m}}^\prime}(\varrho),
\label{FRC}
\end{equation}
preserves the trace so that
\begin{eqnarray*}
&&\sum_{\mathbf{m},\mathbf{m}^\prime}\Tr\left[S^{\mathbf{m},\mathbf{m}^\prime}(\sum_{\vec{m}_{N},\vec{m}_{N}^\prime}D^{\vec{m}_{N},\vec{m}_{N}^\prime}(\tilde{\rho})])\right] \\
&=& \sum_{\vec{m}_{N},\vec{m}_{N}^\prime}\Tr\left[D^{\vec{m}_{N},\vec{m}_{N}^\prime}(\tilde{\rho})]\right].
\end{eqnarray*} 
Now the difference $\Delta W = \langle W\rangle_{N+1} - \langle W\rangle_{N} $becomes
\begin{eqnarray*}
    \Delta W 
    &=& \sum_{\mathbf{m},\mathbf{m}^\prime}\Tr\left[S^{\mathbf{m},\mathbf{m}^\prime}(\sum_{\vec{m}_{N},\vec{m}_{N}^\prime}D^{\vec{m}_{N},\vec{m}_{N}^\prime}(\tilde{\rho})])\right] \nonumber \\
    &&\int dW\,W\left(g_{\Sigma^{2}}(W-\mathcal{W}^{\mathbf{m},\mathbf{m}^\prime}) - g_{\Sigma^{2}}(W)\right)\\
    &=& \sum_{\mathbf{m},\mathbf{m}^\prime}\Tr\left[S^{\mathbf{m},\mathbf{m}^\prime}(\sum_{\vec{m}_{N},\vec{m}_{N}^\prime}D^{\vec{m}_{N},\vec{m}_{N}^\prime}(\tilde{\rho})])\right]\mathcal{W}^{\mathbf{m},\mathbf{m}^\prime},
\end{eqnarray*}
where we shifted the integrand by the amount of $\mathcal{W}^{\vec{m},\vec{m}^\prime}$ and used the  normalization of the Gaussian distributions. On the other hand,
\begin{widetext}
\begin{eqnarray*}
    \langle{W}^{(N+1)}\rangle &=& \int dW^{(N+1)}\,W^{(N+1)}\tilde{P}^{\mathrm{RC}}(W^{(N+1)})\\
    &=& \int dW^{(N+1)}\,\int d\mathbf{x}\,W^{(N+1)}\delta\left(W^{(N+1)}-\sum_{k=4N+1}^{4(N+1)}(-1)^{k}x_{k}\right)\sum_{\vec{m}_{N+1},\vec{m}_{N+1}^\prime}\Tr[D^{\vec{m}_{N+1},\vec{m}_{N+1}^\prime}(\tilde{\rho})] \nonumber \\
    && \times \prod_{k=4N+1}^{4(N+1)}g_{\Sigma^2}\left(x_{k}-\frac{1}{2}(e_{k}^{m_k}+e_{k}^{m_k^\prime})\right).
\end{eqnarray*}
\end{widetext}
Integrating $\mathbf{x}$ first, we get
\begin{widetext}
\begin{eqnarray*}
    \langle{W}^{(N+1)}\rangle &=& \int dW^{(N+1)}\,W^{(N+1)}\sum_{\mathbf{m},\mathbf{m}^{\prime}}\Tr\left[S^{\mathbf{m},\mathbf{m}^\prime}(\sum_{\vec{m}_{N},\vec{m}_{N}^\prime}D^{\vec{m}_{N},\vec{m}_{N}^\prime}(\tilde{\rho})])\right]g(W^{(N+1)}-\mathcal{W}^{\mathbf{m},\mathbf{m}^\prime})\\
    &=& \sum_{\mathbf{m},\mathbf{m}^{\prime}}\Tr\left[S^{\mathbf{m},\mathbf{m}^\prime}\left(\sum_{\vec{m}_{N},\vec{m}_{N}^\prime}D^{\vec{m}_{N},\vec{m}_{N}^\prime}(\tilde{\rho})]\right)\right]\mathcal{W}^{\mathbf{m},\mathbf{m}^\prime} = \Delta W.
\end{eqnarray*}
\end{widetext}

Furthermore, using a fictitious non-normalised density matrix, similar to Eq.~(\ref{eq:B1}),
\begin{eqnarray}
\tilde{\phi}^{\mathrm{RC}}_{\vec{x}_{N+1}}(\tilde{\rho}) &=& \sum_{\vec{m}_{N+1},\vec{m}_{N+1}^\prime} D^{\vec{m}_{N+1},\vec{m}_{N+1}^\prime}(\tilde{\rho}) \nonumber \\
&&\prod_{k=1}^{4(N+1)}\tilde{\sigma}\left(x_k-e_k^{m_k},x_k-e_k^{m_k^\prime}\right),
\end{eqnarray}
with the fictitious pointer state
\begin{equation}
\tilde{\sigma}\left(x_k-e_k^{m_k},x_k-e_k^{m_k^\prime}\right) = g_{\Sigma^2}\left(x_k -\frac{e_k^{m_k}+e_k^{m_k^\prime}}{2}\right),
\end{equation}
we can replicate the proof presented for repeated measurements in~Append.~\ref{Append_RedMapRM}, Eqs.~(\ref{NcycWPDFRM})--(\ref{DRM}) leading to the conclusion that the fictitious energy PDF $\tilde{p}(\vec{x}^N)$ [Eq.~(\ref{pxN})] entering the fictitious work PDF $\tilde{P}^{RC}(W^{(N+1)})$ [Eq.~(\ref{PRCWN})] is determined by the invariant state $\rho^*$ of the map $F^{RC}$, i.e. by the solution of the equation
\begin{equation}
F^{RC}(\rho^*) = \rho^*,
\label{Fr}
\end{equation}
where $F^{RC}$ is defined in Eq.~(\ref{FRC}). Comparing the above equation with Eq.~(\ref{DRM}) we notice the missing exponential terms in the map. This is simply because the exponential terms in case of repeated contacts get simplified, see Eq.~(\ref{RC_W}), and they only affect the initial state of the engine. In other words, in case of repeated measurements to obtain the state of the engine after $N$ cycles we apply the map $F^{\mathrm{RM}}$ $N$ times on the initial density matrix $\rho$ whereas in case of repeated contacts the map $F^{\mathrm{RC}}$ is applied $N$ times on the fictitious initial density matrix $\tilde{\rho}$.

As trace and positivity preserving maps, $F^{RC}$ and $F^{RM}$ possess  at least one invariant solution. The uniqueness of the invariant solution is guaranteed by the dissipative nature of the thermalization strokes. As a conseqence, the invariant densities can be iteratively reached as $\rho^* = \lim_{N\to \infty} (F^{X})^N(\rho)$, where $X = RC, RM$. Accordingly, the spectra of these maps contain one non-degenerate eigenvalue $\Lambda_1 =1$ and the absolute values of all other eigenvales are smaller than one.

In Fig.~\ref{Fig5}{\bf b}-{\bf c}, the second-largest eigenvalues $\Lambda_{2}$ of the maps $F^{\mathrm{RC(RM)}}$ are displayed as  functions of the durations $T_{1}$ and $T_{2}$. When $\Lambda_{2}$ is close to 1, the engine needs a long time to converge to the asymptotic cycle and vice versa.

\subsection{Simplification due to dynamical decoupling}
\label{Append_DDsimp}
In order to prove Eq.~(\ref{RC_simplifying_DD}) we consider the coefficient,
\begin{widetext}
\begin{equation}
D^{\vec{m},\vec{m}^\prime} = \Tr[\mathcal{P}^{m_{4N}}_{4N} \cdots \mathcal{P}^{m_{2i+1}}_{2i+1}\Phi_{u}\left( \mathcal{P}^{m_{2i}}_{2i} \cdots \rho \cdots \mathcal{P}^{m_{2i}^\prime}_{2i}\right)\,\mathcal{P}^{m_{2i+1}^\prime}_{2i+1}\cdots\mathcal{P}^{m_{4N}^\prime}_{4N}],
\label{DcoeffRC}
\end{equation}
\end{widetext}
and focus on the terms involving $2i+1$ and $2i$ subscripts. Now,two alternative cases exist:
\begin{enumerate}
\item $m_{2i} = m_{2i}^{\prime}$ and 
\item $m_{2i} \neq m_{2i}^{\prime}$,
\end{enumerate}
for which we study the central term in Eq.~(\ref{DcoeffRC}) within the trace, i.e.,
\begin{equation}
\langle m_{2i+1}|\Phi_{u}\left( \mathcal{P}^{m_{2i}}_{2i} \cdots \rho \cdots \mathcal{P}^{m_{2i}^\prime}_{2i}\right) |m_{2i+1}^\prime\rangle \propto A,
\end{equation} 
with $A=\langle m_{2i+1}|\Phi_u\left(|m_{2i}\rangle\langle m_{2i}^\prime|\right)|m_{2i+1}^\prime\rangle$. 

For case (i), we could have either $m_{2i+1} = m_{2i+1}^\prime$ or $m_{2i+1} \neq m_{2i+1}^\prime$. For $m_{2i+1} = m_{2i+1}^\prime$, Eq.~(\ref{RC_simplifying_DD}) is satisfied. Whereas for $m_{2i+1} \neq m_{2i+1}^\prime$, $A = \Tr [L^{m_{2i+1}}_u\Phi_u(\mathcal{P}^{m_{2i}}_{2i})]$ which is zero due to Eq.~(\ref{DD}). Hence, for case (i) only when Eq.~(\ref{RC_simplifying_DD}) is satisfied we have a contribution to the work distributions.

For case (ii), we divide into three sub-cases: (iia) $m_{2i+1} = m_{2i+1}^\prime$ for which $A = \Tr [\mathcal{P}^{m_{2i+1}}_{2i+1}\Phi_u(L_u^{m_{2i}})]$ is again zero due to Eq.~(\ref{DD}), (iib) $m_{2i} = m_{2i+1}^\prime$ and $m_{2i}^\prime = m_{2i+1}$ for which $A = \Tr[L_u^{\pm}\Phi_u(L_u^{\mp})] = 0$, and (iic) $m_{2i} = m_{2i+1}$ and $m_{2i}^\prime = m_{2i+1}^\prime$ for which Eq.~(\ref{RC_simplifying_DD}) is satisfied. All other cases like $m_{2i} = m_{2i+1}$ and $m_{2i}^\prime \neq m_{2i+1}^\prime$ reduce to case (iia) since we deal with a two-level working substance. 

This exhausts all possible cases and overall only when Eq.~(\ref{RC_simplifying_DD}) is satisfied we get a contribution to the work distribution.

\section{Reduction in computational costs}\label{Append_Compred}
In this section, we device a computational scheme to reduce the complexity from exponential ($d^{8N}$ with $d$ being the dimension of working substance Hilbert space and $N$ being the number of cycles) to quadratic ($N^2$) as proposed in the main text. Our proof below focuses on the work PDFs given by Eqs.~(\ref{PRMW}) and (\ref{PRC1P}),
\begin{widetext}
\begin{eqnarray}
P^{\mathrm{RM}}(W) &=& \sum_{\vec{m}_N,\vec{m}^{\prime}_N} D^{\vec{m}_N,\vec{m}^{\prime}_N} e^{-\frac{1}{8 \Sigma^2}{\sum_{k=1}^{4N}} (e^{m_k}_k-e^{m_k\prime}_k)^2 } g_{4N\Sigma^2}(W-\mathcal{W}^{\vec{m}_N,\vec{m}^{\prime}_N}), \nonumber \\
 P^{\mathrm{RC}}(W) &=& 
 \sum_{\vec{m}_N,\vec{m}^{\prime}_N} D^{\vec{m}_N,\vec{m}^{\prime}_N}  e^{-\frac{1}{8 \Sigma^2} (\mathcal{W}^{\vec{m}_N} - \mathcal{W}^{\vec{m}{\prime}_N})^2}  g_{\Sigma^2}(W-\mathcal{W}^{\vec{m}_N,\vec{m}^{\prime}_N}).
 \label{PRMRCwN}
\end{eqnarray}
\end{widetext}
Above we have reintroduced the subscript $N$ indicating the cycle number which was suppressed in the main text. The scheme for the heat PDFs can be derived in exactly the same way. The above equations can be rewritten as,
\begin{equation}
P^{\mathrm{X}}(W) = \sum_{\mathcal{W}_N} D^{\mathrm{X}}_{\mathcal{W}_N}\: g_{\Sigma^2_X}(W-\mathcal{W}_N),
\label{PRMX}
\end{equation}
where $\mathrm{X} = \mathrm{RM},\mathrm{RC}$ stands for repeated measurements or repeated contacts. The variances of the superimposed Gaussians depend  on the type of monitoring  with $\Sigma^2_{RM} = 4 N\Sigma^2$ and $\Sigma^2_{RC} = \Sigma^2$. The sum runs over all possible values of $\mathcal{W}_N =\mathcal{W}^{\vec{m}_N,\vec{m}^{\prime}_N}$ and the coefficients $D^{\mathrm{X}}_{\mathcal{W}_N}$ are given by
\begin{equation}
D^{\mathrm{X}}_{\mathcal{W}_N} = \Tr  [\mathcal{D}^{\mathrm{X}}_{\mathcal{W}_N}(\rho)].
\label{DXRM}
\end{equation} 
The operators $\mathcal{D}^{\mathrm{X}}_{\mathcal{W}_N}(\rho)$ are defined as
\begin{eqnarray}
\mathcal{D}^{\mathrm{RM}}_{\mathcal{W}_N}(\rho)  &=&\mkern-36mu \sum_{\stackrel{\vec{m}_N,\vec{m}^{\prime}_N}{\mathcal{W}_N=\mathcal{W}^{\vec{m}_N,\vec{m}^{\prime}_N}}} \mkern-32mu D^{\vec{m}_N,\vec{m}^{\prime}_N}(\rho) e^{-\frac{1}{8 \Sigma^2}\sum_{k=1}^{4N} \left (e^{m_k}_k - e^{m^\prime_k}_k\right )^2}, \nonumber \\
\mathcal{D}^{\mathrm{RC}}_{\mathcal{W}_N}(\rho)&=&\mkern-36mu \sum_{\stackrel{\vec{m}_N,\vec{m}^{\prime}_N}{\mathcal{W}_N=\mathcal{W}^{\vec{m}_N,\vec{m}^{\prime}_N}}} \mkern-32mu D^{\vec{m}_N,\vec{m}^{\prime}_N}(\rho) e^{-\frac{1}{8 \Sigma^2}\left (e^{m_1}_1 - e^{m^\prime_1}_1\right )^2}.
\label{DRCr}
\end{eqnarray}
In case of repeated contacts, the general expression reduces to the above due to the dynamical decoupling between populations and coherence [see Append.~\ref{Append_RedMapRC} Eq.~(\ref{RC_W})]. The reduction in computational resources occurs because we only need to compute the coefficients for a fixed $\mathcal{W}_N$ that scale as $N^2$ rather than the exponential scaling of $d^{8N}$.

In order to see this reduction in computational resources, let us apply the map $S^{\mathbf{m},\mathbf{m}^{\prime}}$, defined in Eq.~(\ref{Smap}), to both sides of Eq.~(\ref{DRCr}), giving,
\begin{eqnarray}
S^{\mathbf{m},\mathbf{m}^{\prime}}\left(\mathcal{D}^{\mathrm{RC}}_{\mathcal{W}_N}(\rho)\right)&=& \mkern-36mu \sum_{\stackrel{\vec{m}_N,\vec{m}^{\prime}_N}{\mathcal{W}_N=\mathcal{W}^{\vec{m}_N,\vec{m}^{\prime}_N}}}\mkern-32mu S^{\mathbf{m},\mathbf{m}^{\prime}} \left(D^{\vec{m}_N,\vec{m}^{\prime}_N}(\tilde{\rho})\right).
\end{eqnarray}
Here we derive the scaling reduction for the case of repeated contacts and the proof for repeated measurements can be obtained in the same spirit. The exponential terms appearing in Eq.~(\ref{DRCr}) have been absorbed in the effective initial density matrix $\tilde{\rho}$ [see Eq.~(\ref{rhods})] since they only depend on the first contact ($m_1$ and $m_1'$), hence, only alters the initial state. It is important to note here that our proof works only if the exponentials appearing in Eq.~(\ref{PRMRCwN}) can be expressed as a product of exponentials implying that such a dramatic reduction is always possible for repeated measurements and is impossible if the populations and coherence do not dynamically decouple in the case of repeated contacts. Using Eq.~(\ref{Dmap}) on the r.h.s. we obtain,
\begin{eqnarray}
S^{\mathbf{m},\mathbf{m}^{\prime}}\left(\mathcal{D}^{\mathrm{RC}}_{\mathcal{W}_N}(\rho)\right)&=& \mkern-36mu \sum_{\stackrel{\vec{m}_N,\vec{m}^{\prime}_N}{\mathcal{W}_N=\mathcal{W}^{\vec{m}_N,\vec{m}^{\prime}_N}}} \mkern-32mu D^{\vec{m}_{N+1},\vec{m}^{\prime}_{N+1}}(\tilde{\rho}).
\label{Compred}
\end{eqnarray}
Recall that $\mathbf{m}^{(\prime)}$ arises from the sequence $\vec{m}^{(\prime)}_{N+1} = \vec{m}^{(\prime)}\oplus \mathbf{m}^{(\prime)}$, defined below Eq.~(\ref{Dmap}). Summing both sides of Eq.~(\ref{Compred}) over $\mathbf{m}^{(\prime)}$ and $\mathcal{W}_N$ (constrained by $\mathcal{W}_{N+1} = \mathcal{W}^{\vec{m}_{N+1},\vec{m}^{\prime}_{N+1}} = \mathcal{W}_N + \mathcal{W}^{\mathbf{m},\mathbf{m}^\prime}$) we obtain,
\begin{eqnarray}
\sum_{\stackrel{\mathbf{m},\mathbf{m}^{\prime},\mathcal{W}_N}{\mathcal{W}_{N+1} = \mathcal{W}_N + \mathcal{W}^{\mathbf{m},\mathbf{m}^\prime}}}\mkern-42mu S^{\mathbf{m},\mathbf{m}^{\prime}}\left(\mathcal{D}^{\mathrm{RC}}_{\mathcal{W}_N}(\rho)\right)
&=&\mkern-54mu \sum_{\stackrel{\vec{m}_{N+1},\vec{m}^{\prime}_{N+1}}{\mathcal{W}_{N+1}=\mathcal{W}^{\vec{m}_{N+1},\vec{m}^{\prime}_{N+1}}}} \mkern-54mu D^{\vec{m}_{N+1},\vec{m}^{\prime}_{N+1}}(\tilde{\rho}) \nonumber \\
&=& \mathcal{D}^{\mathrm{RC}}_{\mathcal{W}_{N+1}}(\rho).
\end{eqnarray}
Therefore, the operator $\mathcal{D}^{\mathrm{RC}}_{\mathcal{W}_{N+1}}(\rho)$ is obtained recursively from $\mathcal{D}^{\mathrm{RC}}_{\mathcal{W}_N}(\rho)$ in terms of a sum over $\mathbf{m}$, $\mathbf{m}^{\prime}$ and $\mathcal{W}_N$. The required number of computations does not scale with $N$ (for $\mathbf{m}^{(\prime)}$ we have $2^4$ terms for $d=2$) and the sum over $\mathcal{W}_N$ scales as $N^2$. Thus, the computations performed, due to the grouping with fixed work values, reduces from an exponential scaling of $2^{8N}$ ($d=2$) to a quadratic $N^2$.

\section{Equivalence between one- and two-pointer work PDFs for a perfectly thermalizing adiabatic engine}\label{Append_Equiv}
In this section, we show that for a perfectly thermalizing adiabatic engine diagnosed via repeated contacts the one- and two-pointer work PDFs are equal independent of the dimensionality $d$ of the Hilbert space of the working substance. In order to achieve our objective, we begin with the one- and two-pointer work PDFs as defined in Eqs.~(\ref{PRC2P}) and (\ref{PRC1P}), i.e.,
\begin{widetext}
\begin{eqnarray}
 P^{\mathrm{RC,1P}}(W) &= &
 \sum_{\vec{m},\vec{m}^\prime} D^{\vec{m},\vec{m}^\prime}  e^{-\frac{1}{8 \Sigma^2} (\mathcal{W}^{\vec{m}} - \mathcal{W}^{\vec{m}\prime})^2}  g_{\Sigma^2}(W-\mathcal{W}^{\vec{m},\vec{m}^\prime}), \nonumber \\
 P^{\mathrm{RC,2P}}(W)
&= & \sum_{\vec{m},\vec{m}^\prime} D^{\vec{m},\vec{m}^\prime}  e^{-\frac{1}{8 \Sigma^2} (\mathcal{W}^{\vec{m}} - \mathcal{W}^{\vec{m}\prime})^2} e^{-\frac{1}{8 \Sigma^2} (\mathcal{Q}^{\vec{m}} - \mathcal{Q}^{\vec{m}\prime})^2} g_{\Sigma^2}(W-\mathcal{W}^{\vec{m},\vec{m}^\prime}). \nonumber
\end{eqnarray}
\end{widetext}
Comparing the above equations, it is easy to see that the PDFs are equal when either $\mathcal{Q}^{\vec{m}} = \mathcal{Q}^{\vec{m}\prime}$ or when $D^{\vec{m},\vec{m}^\prime} = 0$. 

We restrict ourselves to a single cycle ($N=1$) perfectly thermalizing engine, such that the elements of $\vec{m}^{(\prime)}=(m_1^{(\prime)},m_2^{(\prime)},m_3^{(\prime)},m_4^{(\prime)})$. To proceed, we consider all possible relations between the elements of $\vec{m}$ and $\vec{m}^{\prime}$, i.e.,  
\begin{enumerate}
\item $m_2 = m_2'$ and $m_3 = m_3'$ (independent of $m_1^{(\prime)}$ and $m_4^{(\prime)}$),
\item $m_2 \neq m_2'$ and $m_3 = m_3'$ (independent of $m_1^{(\prime)}$ and $m_4^{(\prime)}$), and
\item $m_3 \neq m_3^\prime$ (independent of $m_1^{(\prime)}$, $m_2^{(\prime)}$, and $m_4^{(\prime)}$),
\end{enumerate}
and show that they can be classified into one of the two categories ($\mathcal{Q}^{\vec{m}} = \mathcal{Q}^{\vec{m}\prime}$ or $D^{\vec{m},\vec{m}^\prime} = 0$).\\
~\\
\noindent {\bf Case (i) (\emph{Equality due to matching heat outcomes}):} For a single cycle considered herein the heat outcomes $\mathcal{Q}^{\vec{m}} = e^{m_3}_3 - e^{m_2}_2$ and $\mathcal{Q}^{\vec{m}'} = e^{m_3'}_3 - e^{m_2'}_2$, using Eq.~(\ref{QRC}). Since $m_2 = m_2'$ and $m_3 = m_3'$, clearly $\mathcal{Q}^{\vec{m}} = \mathcal{Q}^{\vec{m}\prime}$.\\
~\\
\noindent {\bf Case (ii) (\emph{Equality due to perfect thermalization}):} For a single cycle engine, the explicit expression for the coefficient $D^{\vec{m},\vec{m}^\prime}$ reads,
\begin{widetext}
\begin{eqnarray}
D^{\vec{m},\vec{m}^\prime} = \Tr[\mathcal{P}^{m_4}_4 \tilde{U} \mathcal{P}^{m_3}_3 \, \Phi_{h}\left( \mathcal{P}^{m_2}_2 U \mathcal{P}^{m_1}_1 \rho \mathcal{P}^{m_1'}_1 U^{\dagger} \mathcal{P}^{m_2'}_2\right)\,\mathcal{P}^{m_3'}_3\tilde{U}^{\dagger}\mathcal{P}^{m_4'}_4],
\label{DmmappendC}
\end{eqnarray}
\end{widetext}
with the operators $\mathcal{P}^{m_j}_j$ projecting into the eigenstates of Hamiltonian $H_j$, $\rho$ being the density matrix of the working substance, and $U$ ($\tilde{U}$) being the unitary forward (reversed) time-evolution operators. Since the engine is perfectly thermalizing, $\Phi_h(\rho) = \rho_h \Tr[\rho]$ using Eq.~(\ref{E:1}), we obtain
\begin{widetext}
\begin{eqnarray}
\Phi_{h}\left( \mathcal{P}^{m_2}_2 U \mathcal{P}^{m_1}_1 \rho \mathcal{P}^{m_1'}_1 U^{\dagger} \mathcal{P}^{m_2'}_2\right) = \rho_h \Tr[\mathcal{P}^{m_2}_2 U \mathcal{P}^{m_1}_1 \rho \mathcal{P}^{m_1'}_1 U^{\dagger} \mathcal{P}^{m_2'}_2] = 0.
\end{eqnarray}
\end{widetext}
The rightmost equality is because we consider $m_2 \neq m_2'$ in this case giving us the trace in the middle term to be zero. Therefore, for case (ii) $D^{\vec{m},\vec{m}^\prime} = 0$.\\
~\\
\noindent{\bf Case (iii) (\emph{Equality due to adiabaticity}):} We further sub-divide case (iii) into two sub-categories: (a) $m_3 \neq m_4$ or $m_3' \neq m_4'$ and (b) $m_3 = m_4$ and $m_3' = m_4'$. 

For case (iiia), since the evolution during the work strokes is adiabatic, i.e., if the system starts in an eigenstate of the initial Hamiltonian, it will end in the corresponding eigenstate of the final Hamiltonian, either 
\begin{eqnarray}
\mathcal{P}^{m_4}_4 \tilde{U} \mathcal{P}^{m_3}_3 &=& 0\quad \mathrm{when~} m_3 \neq m_4~~\mathrm{or} \nonumber \\
\mathcal{P}^{m_3'}_3\tilde{U}^{\dagger}\mathcal{P}^{m_4'}_4 &=& 0 \quad \mathrm{when~} m_3'\neq m_4'.
\end{eqnarray}
In either case the coefficient $D^{\vec{m},\vec{m}^\prime} = 0$ [see Eq.~(\ref{DmmappendC})].

For the case (iiib), since $m_3\neq m_3'$, $m_3 = m_4$, and $m_3' = m_4'$ implies $m_4 \neq m_4'$. Therefore, again $D^{\vec{m},\vec{m}^\prime} = 0$ due to the overall trace in Eq.~(\ref{DmmappendC}). Thus, overall for case (iii) $D^{\vec{m},\vec{m}^\prime} = 0$.

Overall, with the three cases above, all possible combinations of $\vec{m}$ and $\vec{m}'$ are excluded that would lead to a difference between the one- and the two-pointer work PDFs. Moreover, our proof does not depend on the working substance of the engine and hence holds for all thermalizing adiabatic quantum Otto engines diagnosed via repeated contacts.
\bibliography{Son_References}

\end{document}